\documentclass[aps,twocolumn,showlabels,showrefs,amsmath,amssymb,prl,superscriptaddress,floatfix,colors]{revtex4-2}

\usepackage{lineno}
\usepackage{graphicx}
\usepackage{dcolumn}
\usepackage{bm}
\usepackage{cancel}
\usepackage{graphicx}
\usepackage{dcolumn}
\usepackage{bm}
\usepackage{amssymb}
\usepackage{hyperref}

\usepackage{multirow}
\usepackage{color}
\usepackage[normalem]{ulem}
\newcommand{\ylann}[1]{{\color{black} #1}}

\usepackage[cp1251]{inputenc}

\makeatletter
\newcommand*{\sumcirclearrowleft}{%
 \DOTSB
 \mathop{
  \mathchoice
   {\rlap{\kern.25em\rotatebox[origin=c]{-90}{$\circlearrowleft$}}{\sum}}
   {\vcenter{\rlap{\kern.2em\rotatebox[origin=c]{-90}{$\scriptscriptstyle\circlearrowleft$}}}{\sum}}
   {\sum}{\sum}
 }\slimits@
}

\newcommand*{\sumcirclearrowright}{%
 \DOTSB
 \mathop{
  \mathchoice
   {\rlap{\kern.25em\rotatebox[origin=c]{90}{$\circlearrowright$}}{\sum}}
   {\vcenter{\rlap{\kern.2em\rotatebox[origin=c]{90}{$\scriptscriptstyle\circlearrowright$}}}{\sum}}
   {\sum}{\sum}
 }\slimits@
}
\makeatother

\begin{document}

\title{Non-Reciprocal Interactions Reshape Topological Defect Annihilation}

\author{Ylann Rouzaire} \email{rouzaire.ylann@gmail.com}
 \affiliation{Departament de F\'isica de la Materia Condensada, Universitat de Barcelona, Mart\'i i Franqu\`es 1, E08028 Barcelona, Spain}
  \affiliation{UBICS University of Barcelona Institute of Complex Systems , Mart\'i i Franqu\`es 1, E08028 Barcelona, Spain}
 \author{Daniel J.G. Pearce}
 \affiliation{Universit\'e de Gen\`eve D\'{e}partement de physique th\'{e}orique, 24 quai Ernest-Ansermet, 1211 Gen\`{e}ve, Switzerland }
 
 \author{Ignacio Pagonabarraga} 
 \affiliation{Departament de F\'isica de la Materia Condensada, Universitat de Barcelona, Mart\'i i Franqu\`es 1, E08028 Barcelona, Spain}
  \affiliation{UBICS University of Barcelona Institute of Complex Systems , Mart\'i i Franqu\`es 1, E08028 Barcelona, Spain}

 \author{Demian Levis} 
 \affiliation{Departament de F\'isica de la Materia Condensada, Universitat de Barcelona, Mart\'i i Franqu\`es 1, E08028 Barcelona, Spain}
 \affiliation{UBICS University of Barcelona Institute of Complex Systems , Mart\'i i Franqu\`es 1, E08028 Barcelona, Spain}

\date{\today}
\begin{abstract}
   We show how non-reciprocal ferromagnetic interactions between neighbouring planar spins in two dimensions, affect the behaviour of topological defects. Non-reciprocity is introduced by weighting the coupling strength of the two-dimensional XY model by an anisotropic kernel. As a consequence, in addition to the topological charge $q$, the actual shape of the defects becomes crucial to faithfully describe their dynamics. Non-reciprocal coupling twists the spin field, selecting  specific defect shapes, dramatically altering the pair annihilation process. Defect annihilation can either be enhanced or hindered, depending on the shape of the defects concerned and the degree of non-reciprocity in the system. We introduce a continuous description -- for which the phenomenological coefficients can be explicitly written in terms of the microscopic ones -- that captures the  behaviour of the lattice model.  
\end{abstract}
\maketitle

Collective excitations in the form of topological defects are known to control the large-scale behaviour of a broad class of equilibrium systems, and to be eventually responsible for a Kosterlitz-Thouless (KT) phase transition, as exhibited by the two-dimensional ($2d$) XY model \cite{Berezinskii1971, Kosterlitz1973, Kosterlitz1974}. Being local singularities of the relevant order parameter field, topological defects can be considered as quasi-particles (characterised by their charge $q$), robust against perturbations \cite{mermin1979topological, chaikin1995principles, nelson2002defects}, and the study of their dynamics provides insights into the ordering mechanisms at play \cite{kibble1976topology, zurek1985cosmological,  YurkeHuse1993, JelicCugliandolo, pearce_properties_2021}. 
Out-of-equilibrium, topological defects exhibit distinctive dynamical features, such as self-propulsion or super-diffusion, often breaking the symmetry between oppositely charged defects \cite{giomi2014defect, pearce2020defect, vitelli_defects_odd_elasticity, RouzaireLevis, rouzairedynamics, shankar2022topological, bililign2022motile, ChardacBartolo2021, poncet_when_2022}. 
{The study of dynamics of defects in active matter is currently providing new light  into the physics of several biological processes}
 \cite{saw2017topological, kawaguchi2017topological, copenhagen2021topological, maroudas2021topological}. Although the impact of defects on the critical properties of active matter is still to be settled \cite{LinoDefects, shankar2018defect, pearce2021orientational, shi2023extreme}, understanding their dynamics can clarify self-organisation mechanisms in living systems and eventually inspire the design of novel materials. 

Non-reciprocal (NR) interactions provide an alternative route to conceive systems locally driven out-of-equilibrium. 
Indeed, the effective interactions between non-equilibrium agents  generically violates the \emph{actio-reactio} principle. Examples include microorganisms \cite{agudo-canalejo_active_2019}, 
 social agents {(decision making, epidemic spreading, etc.)\cite{abrams2000evolution, couzin2005effective, castellano2009statistical, pastor2015epidemic},  ecological communities \cite{ros2023generalized}}, catalytic colloids \cite{grauer2021active} or specifically tailored robot swarms \cite{rubenstein2014programmable, fruchart_non-reciprocal_2021, march2023honeybee}.
The analysis of both particle and spin models, together with  continuum theories, have recently shown that non-reciprocity leads to distinct emergent phenomena 
 \cite{ivlev_statistical_2015, dadhichi_nonmutual_2020, you2020nonreciprocity, saha_scalar_2020, fruchart_non-reciprocal_2021, poncet_when_2022, han2021fluctuating, avni2023non, benois2023enhanced, rana2023defect, hickey2023nonreciprocal, loos2023long,seara2023non, vafa2022defect, besse2022metastability}. 
{ In particular, continuous theories of constant-density flocks \emph{\`a la }Toner-Tu, should capture the large scale behaviour  of non-reciprocal continuous spins \cite{dadhichi_nonmutual_2020, solonJclub, besse2022metastability}}.

In this manuscript we uncover the effect of NR interactions on topological defects. We start from an extension of the $2d$XY model where a given spin interacts with its neighbourhood in a way that depends on the {individual} state of the spin itself {in a non-reciprocal way}: interactions are stronger along the spins' heading direction {regardless of the orientation of its neighbours}. We show that NR interactions affect the shape of $+1$ isolated defects, which in turn impacts their annihilation kinetics and coarsening dynamics. We also introduce a continuum description which reproduces these results and gives 
 a microscopic origin to the phenomenological coefficients in the Toner-Tu equation.



\begin{figure}[h!]
    \centering
    \includegraphics[width=1\linewidth]{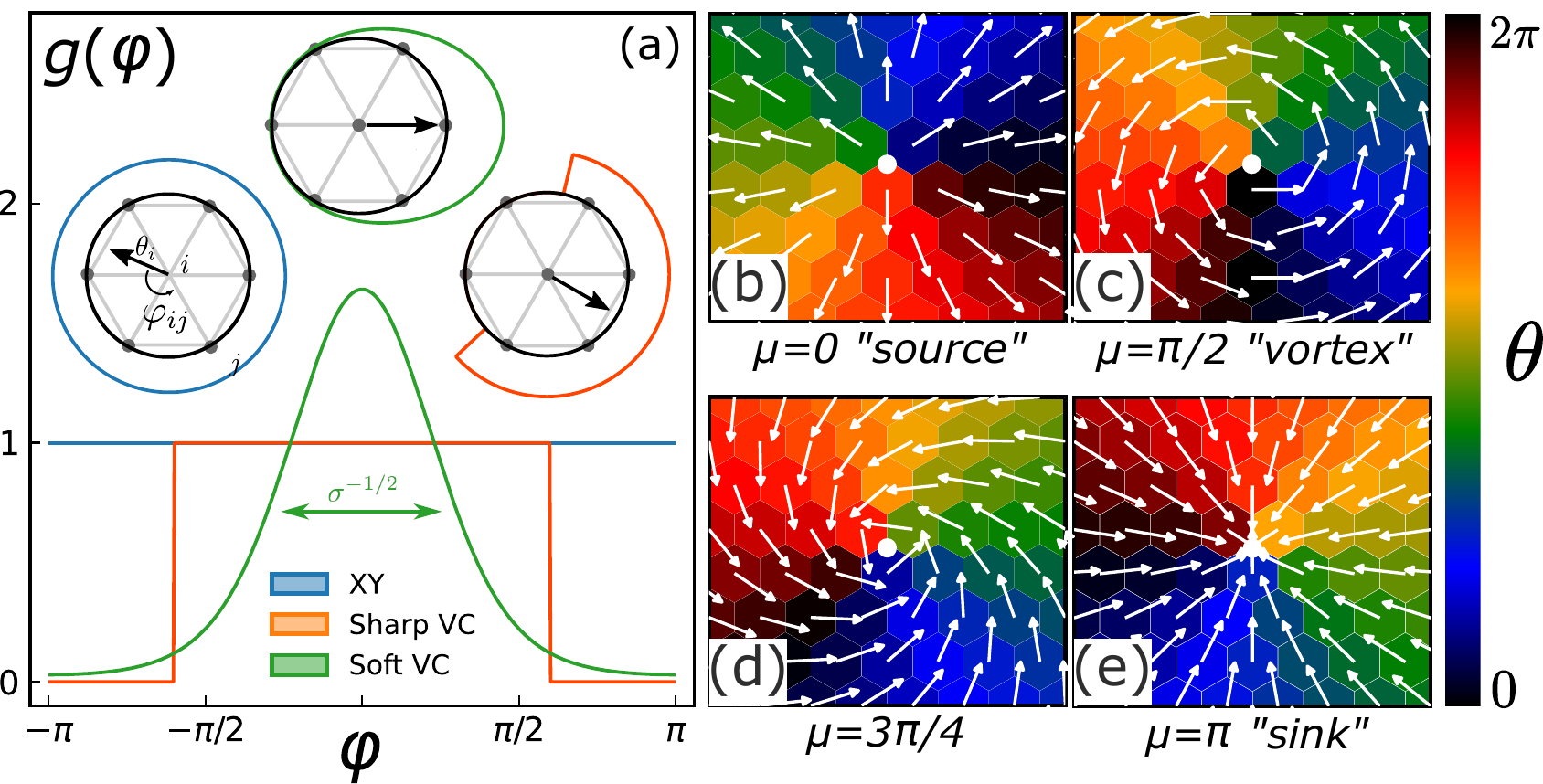}   
    \caption{\textbf{(a)} Sketch of the nearest neighbours coupling: the isotropic reciprocal case, in blue, the sharp vision cone, in orange, and our smooth version, in green (using $\sigma = 5$).  \textbf{(b-e)} Four +1 defects with different shapes $\mu_+$ .
     White dots show  the defect core, on the dual (hexagonal) lattice. The colour code displays the phase $\theta$ of each spin.
    }
    \label{fig:1}
\end{figure}

\begin{figure*}
    \centering
    \includegraphics[width=0.99\linewidth]{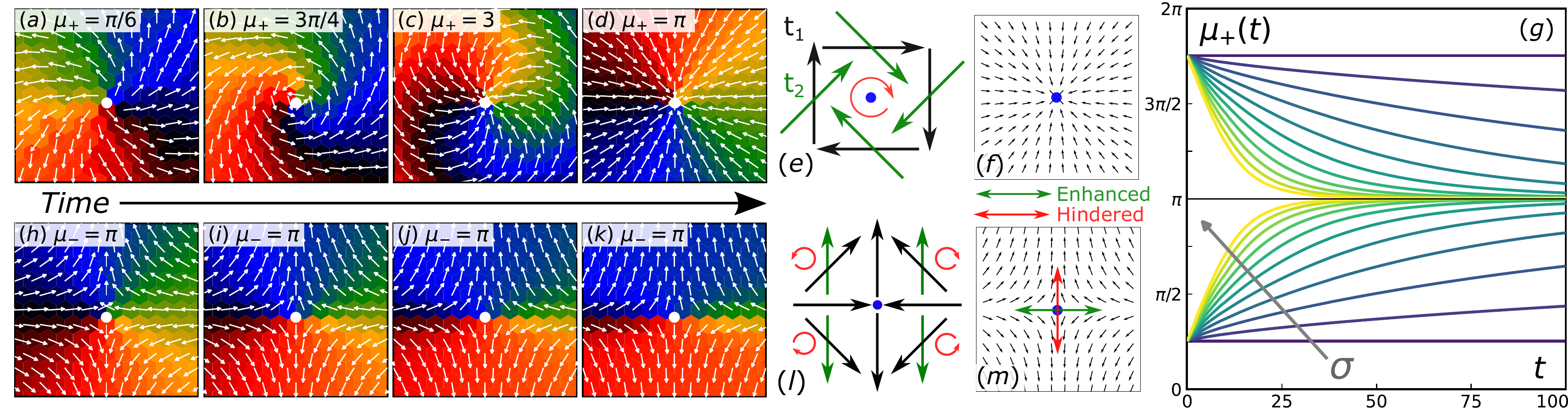} 
    \caption{
    \textbf{(a-d)} Twist of a positive defect towards the sink state ($\mu_+ = \pi$). Same color code as in Fig.\ref{fig:1}. 
    \textbf{(h-k)} Polarisation of a negative defect. 
     \textbf{(e,l)} Reshaping mechanism for $+1$  (e) and  $-1$ (l)  defects, at time $t_1$ (black) and $t_2 > t_1$ (green).  The blue circles are the defect cores; the red arrows represent the non-reciprocal torques.  \textbf{(f,m)} Analytical final states computed in SM \cite{SM}, for $+1$ (f) and $-1$ (m) defects. For the $-1$ defect, we indicate with a green (resp. red) arrow the direction parallel (resp. perpendicular) to the defect axis, along which the motion of the defect is enhanced (resp. hindered).   
    \textbf{(g)} Time evolution a $q=+1$ defect's shape for different $\sigma$ from 0 (dark blue) to $0.5$ (yellow) by increments of $0.05$, for two initial configurations ($\mu_0=\pi/4,\,7\pi/4$).
}
    \label{fig:2}
\end{figure*}

The NR $2d$XY model is composed of spins $\boldsymbol{S}_i=(\cos\theta_i,\sin\theta_i)$ sitting on the nodes of a triangular lattice of linear size $L$, and evolving accordingly to  
\begin{equation}
\gamma\,\dot{\theta}_i=J\sum_{j \in \partial_i} \,g_{\sigma}(\varphi_{i j})\sin \left(\theta_j-\theta_i\right) + \sqrt{2\gamma\, T}\,\eta_i(t) 
 \label{eq:eom}
 \end{equation}
 
where $\gamma$ is the damping coefficient, $J$ the coupling, $\eta$ a Gaussian white noise with zero mean and unit variance and $T$ the temperature of the bath to which the system is coupled (fixing $k_B=1$).  
NR interactions are introduced by weighting the coupling between two neighbouring spins, say $i$ and $j$, according to the orientation of spin $i$ with respect to the direction of the bond connecting the two, denoted $\varphi_{ij}$ (see sketch  Fig.\ref{fig:1}(a)). This is encoded in the kernel $g_{\sigma}$: the reciprocal $2d$XY model with overdamped (non-conserved) dynamics \cite{YurkeHuse1993} is recovered for constant $g_{\sigma}$, while ``vision cone" interactions would correspond to a step function kernel \cite{loos2023long, sarahloosXYalignment, couzin2005effective, chen2017fore}. 
We consider instead a ``smooth vision cone", based on the Von Mises distribution: $g_\sigma(\varphi) = \exp(\sigma\,\cos\,\varphi)$ \cite{VonMises} \ylann{in an effort to  
attenuate discretisation effects due to the  lattice geometry 
(cf. \cite{sarahloosXYalignment, popli2025don} and discussion in the SM \cite{SM}).  This approach also allows to explore} other ways of breaking the action-reaction principle. Since in the small $\varphi$ regime  $g_\sigma(\varphi) \sim e^{-\varphi^2 / 2\sigma^{-1}}$,  $\sigma$ plays the role of the inverse variance (higher $\sigma$ values  shrink the vision cone).

The model is thus fully controlled by two non-dimensional parameters: the \emph{reduced temperature} $T\equiv T/J$, 
and the \emph{non-reciprocal parameter} $\sigma\geq 0$.
Time is expressed in units of $\gamma/J$.
Importantly, as $g_{\sigma}(\varphi_{ij})$ depends on $\theta_i$ but not $\theta_j$, Eq.(\ref{eq:eom}) cannot be thought of
as an equilibrium dynamics driven by a Hamiltonian-like function $\mathcal{H}=-J\sum g_{\sigma}(\varphi_{ij}) \,\boldsymbol{S}_i\cdot\boldsymbol{S}_j$. Indeed, in the presence of NR interactions the energy of a bond is ambiguous, as $g_{\sigma}(\varphi_{ij})\neq g_{\sigma}(\varphi_{ji})$. Generally speaking, the defining feature of a NR system is its dynamics, which here we chose to be Langevin-like \cite{RouzaireLevis,fruchart_non-reciprocal_2021} (contrary to Glauber type as in \cite{loos2023long}). 
Thus, $\sigma$ breaks \emph{parity}, \emph{O}(2) \emph{rotational} and \emph{time-reversal} symmetry \cite{loos_irreversibility_2020, suchanek2023time} (see SM\cite{SM}).

The $2d$XY model features point-like topological defects with integer charge, $q$, given by the winding number~\cite{mermin1979topological}. 
The local field around a $q = \pm1$ defect located at the origin is given by \ylann{$\theta_\pm(x,y) = q\,\text{atan}(y/x) + \mu_\pm$}, where the integration constant $ \mu_\pm$ physically represents its \emph{shape} (see Fig.\ref{fig:1}(b-e) for a $q = +1$ defect). Building on this reference form, we define, in our NR model, the  \ylann{scalar (modulo $2\pi$) shape of a $q=\pm 1$ defect as ${\mu_\pm}~=~\text{Arg}(\sum_{j} \exp(i[\theta_j - q\,\text{atan}(y_j/x_j)]) )$}, where the sum runs over the sites defining the plaquette in which the defect lies, \ylann{see Figures~3 and 4 of the SM} \cite{SM}. 
{ For $-1$ defects, $\mu_-$ corresponds to a global rotation of the defect by $\mu_-/2$. However, for $+1$ defects, $\mu_+$ changes the shape of the defect, giving \emph{sources} ($\mu_+ = 0$), \emph{sinks} ($\mu_+ = \pm\pi$) and \emph{vortices} ($\mu_+ = \pm\pi/2$).}

\begin{figure*}
    \centering
    \includegraphics[width=\linewidth]{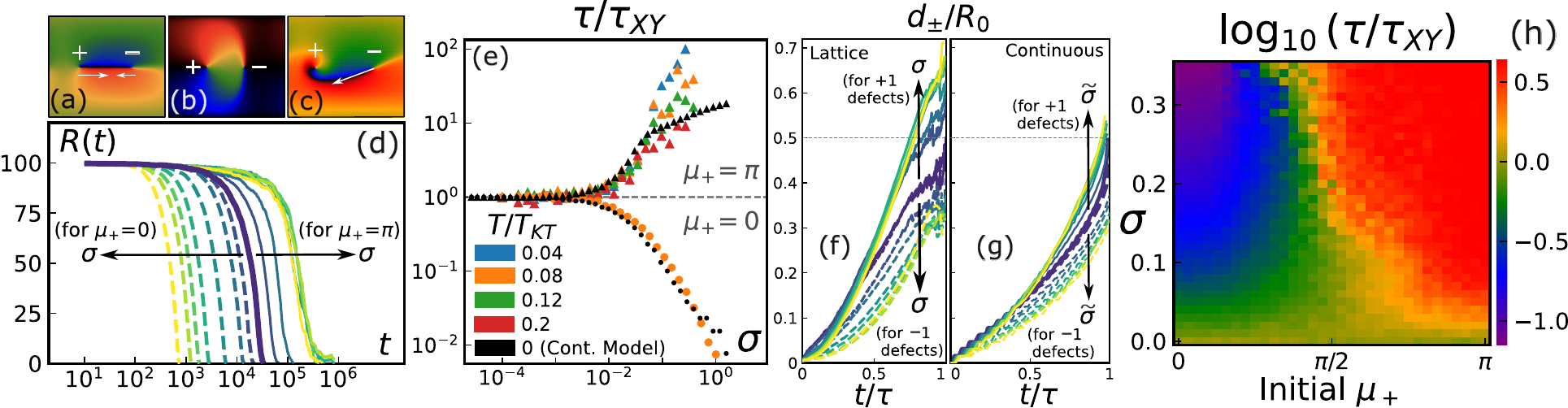}   
    \caption{
     \textbf{(a,b,c)} Configurations of two defects of opposite charge at $T=0, \sigma = 0.35$ with $\mu_+ = 0\,, \pi\,, \pi/2$, respectively. Arrows indicate their velocity. 
     \textbf{(d)} Inter-defect distance $R(t)$ between a  $\mu_+ = 0$ and a $\mu_-=\pi$ (dashed lines) and a $\mu_+ = \pi$ and a $\mu_-=0$ (solid lines) pair, {for $\sigma=0,0.01,0.02, 0.05, 0.1,0.15, 0.2, 0.3$, increasing along the arrows from blue to yellow}. In both cases, $T = 0.08\ T_{KT}, L =2R_0= 200$. 
     \textbf{(e)} Annihilation time $\tau$ rescaled by the one in the $2d$XY case, for {$\mu_+ = 0$ (lower half) and $\mu_+ = \pi$ (upper half)} at different temperatures ($L=2R_0 = 100$). For the continuous model ($L=256$), we plot the data against $2\tilde\sigma/L$ to compare to $\sigma$ (cf. SM \cite{SM}). 
     For $\mu_+=0$, we only display the case $T = 0.08\ T_{KT}$ as curves for different temperatures superimpose for all $\sigma$.
     \textbf{(f)} Distance travelled $d_\pm/R_0$ by  $+1$ (solid lines) and $-1$ (dotted lines) defects in the configuration shown in panel (a) ($\mu_+ = 0$, $\mu_-=\pi$) the same values of $\sigma$ as in panel (d). {In the XY model, both defects travel at the same speed : $d_{\pm}(t=\tau) = R_0/2$ (dashed line).}  \textbf{(g)} {Same quantity for the continuous model, for $\tilde\sigma=0, 0.5, 1, 1.5, 2,2.5,3$.  }
     \textbf{(h)} Colour map of the annihilation time in log-scale 
     as a function of initial shape $0\leq\mu_+\leq\pi$ and non-reciprocity $\sigma$ (using the lattice model, $L=64, R_0 = 30$ and $T = 0.08\ T_{KT}$).
     }
    \label{fig:3}
\end{figure*}

To understand the effect of NR interactions on $\mu_+$ we analyse an isolated $q=+1$ defect. We observe that at zero temperature, the source and sink are stationary states. All intermediate defect shapes decay to a sink, shown for $\mu_+=\pi/6$ in Fig.~\ref{fig:2}(a-d). By approximating the dynamics of $\theta$ close to a defect, we show that $\mu_+=0$ is an unstable fixed point and $\mu_+=\pi$ is stable, see SM~\cite{SM}. To highlight the mechanism behind this effect, consider a $q=+1$ vortex defect ($\mu_+ = \pi/2$). The arrows in front of a given spin point slightly towards the defect core, and the arrows behind  slightly away. In a reciprocal system, these influences balance and the defect shape is stable. NR interactions cause an imbalance between these two influences leading to a torque on the spins, reshaping the defect as sketched in Fig.\ref{fig:2}(e).

To quantify such reshaping, we follow $\mu_{+}(t)$ in time {at $T=0$}. 
As expected, for  $\sigma = 0$, the shape of the defect  remains unchanged, see Fig.\ref{fig:2}(g). For  $\sigma > 0$, $\mu_+$ spontaneously decays to $\pi$ (sink), from any initial value $\mu_0$. 
While $\sigma$ appears to control the decay rate, it does not define a characteristic timescale that can be used to collapse the curves. To gain some insight, we derive the following continuous approximation of Eq.(\ref{eq:eom}) ({on the square lattice for simplicity and transparency},  cf. SM \cite{SM}), at $T=0$, for small $\sigma$, small spatial gradients: 
\begin{equation}
\dot{\theta}=\Delta \theta+2\sigma\, (\nabla \times \boldsymbol{S})_z \equiv f_{\text{EL}} + \sigma \,f_{\text{NR}}
\label{eq:thetas_rot}
\end{equation}
where $\Delta$ is the Laplacian operator and $\boldsymbol{S}$ the spin vector field. From this expression, it is clear that $\sigma$ cannot be absorbed in the time unit and that the dynamics are $\mu_+$-invariant around a $q=1$ defect if and only if $\sigma=0$. 

The reciprocal part of the kernel is responsible for the elastic force $f_{\text{EL}}$ (the $2d$XY model in the spin wave approximation \cite{chaikin1995principles}) 
while the NR part, by explicitly introducing an asymmetry (here front/back), allows spins to be sensitive to the vorticity of the surrounding spin field.
The resulting twisting torque $f_{\text{NR}}$ powers the twist from the inside of the defect's core, where the amplitude of the rotational is greater. The elastic term distributes the stress isotropically, explaining why the twist radially propagates outwards, {see Fig.~\ref{fig:2}(a-d) and Supplementary Movie 1 \cite{SM}. Such twisting is also found in continuum theories of constant-density flocks \cite{vafa2022defect}.}
{Sinks are attractors of the dynamics, while sources are unstable fixed points. 
For $T>0$, the thermal noise induces perturbations with a non-vanishing vorticity which in turn generates a finite $f_\text{NR}$, driving the defect away from the (unstable) source state and towards the sink state.}

We now turn our attention on $q=-1$ defects. Figure~\ref{fig:2}(h) shows a stable $q=-1$ defect in the absence of NR interactions { and Fig.~\ref{fig:2}(h-k) shows its time evolution at $T=0, \sigma>0$.  }
Here it is clear to see that the curl of the spin vector field initially has a clear quadrupole like symmetry, which will be reflected in $f_{\text{NR}}$ (see SM \cite{SM}) and is highlighted in Fig.\ref{fig:2}(l). This results in a stable defect shape with a symmetry axis from which the spin vector points predominantly outwards; see Fig.\ref{fig:2}(k) for a numerical configuration and Fig.\ref{fig:2}(m) for the  final stable shape obtained analytically \cite{SM}.
The existence and relevance at large scales of such structures has been studied by means of continuum models of constant-density flocks in \cite{besse2022metastability}.  
Importantly, the polarised field around a $q=-1$ defect provides a preferred path for defect motion, {indicated in Fig.\ref{fig:2}(m) by green and red arrows, respectively for enhanced and hindered motion}. The location of the defect core can slide along the symmetry axis with a small, continuous variation of the spin vector field. On the contrary, it is very unfavourable for the defect to move perpendicular to the axis as it would require a large number ($\sim L$) of spin flips. 



As NR interactions reshape defects, they strongly impact their annihilation dynamics. 
In the reciprocal $2d$XY model,  defects of opposite charge $\pm q$ at a distance $R$  attract each other with a Coulomb force $F \sim q^2/R$ that drives their annihilation at low  $T<T_{KT}$ \cite{YurkeHuse1993,chaikin1995principles, RouzaireLevis, rouzairedynamics}. 
{On a square lattice, $T_{KT}=0.89$; on a triangular lattice $T_{KT}=1.4$ \cite{SM}. } The XY model is symmetric under parity and thus $\pm q$ defects are equivalent.
NR interactions break such symmetry, therefore can \emph{enhance} or \emph{hinder} the annihilation process and induce effective \emph{transverse forces}, depending on the specific shape of the defects involved. 

To explore the annihilation process, we study a pair of defects of charge $q = \pm 1$ and initial shapes $\mu_\pm$  at a distance $R_0$.
{ We create non-twisted configurations by initially imposing $\mu_+ - \mu_- =  \pi$. As derived in the SM \cite{SM}, this constraint imposes a constant spin orientation along the segment joining the two defect cores.}
We then let the system evolve at $T=0.08\,T_{KT}$, imposing periodic boundary conditions, and track defects over time to obtain their displacement $d_\pm(t)$ and their mutual distance $R(t)$.  
We focus first on two limit situations, involving sources and sinks, that will provide our reference scenarios, to then move to the description of more general annihilation process, as illustrated in Fig.\ref{fig:3}(a-c). 

A source, $\mu_+ = 0$, can only be initially paired with a  $\mu_- = \pi$ defect that creates a symmetry axis along the core-to-core direction, see Fig.\ref{fig:3}(a). Such symmetry provides a preferential annihilation pathway. As shown  in Fig.\ref{fig:3}(d),  $R(t)$ decays faster for larger values of $\sigma$, which translates into a reduced annihilation time $\tau$ (that appears to be independent of $T$, see Fig.\ref{fig:3}(e)). 
Moreover, $q=\pm 1$ defects are no longer equivalent: $+1$ defects move faster than $-1$ defects in the enhanced annihilation process (Fig.\ref{fig:3}(f) {for the lattice model and Fig.\ref{fig:3}(g) for the continuum model introduced later in the text}). {This is in contrast with the equilibrium XY model, for which the invariance under the transformation $\theta \to -\theta$ prevents any differentiation, and with the continuum theory studied by Vafa \cite{vafa2022defect}, where the defect interactions appear to be symmetric. }

For a sink, $\mu_+ = \pi$ ($\mu_-=0$), the symmetry axis of the $-1$ defect grows perpendicular to the core-to-core direction, as illustrated in Fig.\ref{fig:3}(b). {The larger $\sigma$, the faster the axis grows (see \cite{SM}).
Such structure hinders the motion along the core-to-core direction, explaining why $R(t)$ decays slower (solid lines in Fig.\ref{fig:3}(d)): the annihilation is dramatically slowed down, up to a factor $10^2$ with respect to the $2d$XY case, see Fig.\ref{fig:3}(e).}

For any other initial pair, the $+1$ defect experiences a twist due to a non-vanishing $f_{\text{NR}}$ {and its mobility strongly decreases}, while the $-1$ defect grows an oblique symmetry axis that eventually curls to meet the $+1$ defect core, see Fig.~\ref{fig:3}(c). Once the preferential path between the two defects has been created, the $-1$ moves along it until either the defects are close enough to annihilate or the polarisation axis becomes perpendicular to the $+1$/$-1$ direction, {cf. Supplementary Movie 6a}. 
The annihilation time (normalised by $\tau_{XY}, $the corresponding time in the $2d$XY model) for a pair of defects as a function of $\mu_+-$ and $\sigma$ is shown in Fig.\ref{fig:3}(h). As $\sigma$ is increased, annihilation times are either increased for $\mu_+\approx \pi$ (red region) or decreased for $\mu_+$ and large $\sigma$ (blue region in Fig.\ref{fig:3}(h)) with a smooth transition between the two extremes.
Sharp vision cones give qualitatively similar results, see SM \cite{SM}. By imposing the same head-tail asymmetry to both sharp and soft kernels, we obtain an equivalence relation between $\sigma$ and the sharp vision cone aperture $\Theta$, { which leads to a good quantitative agreement between both models, confirming the robustness and generality of our findings \cite{SM}.}

\newpage
At the {coarse-grained} level, our model can be described\,by 
\begin{equation}
\dot{\boldsymbol{S}} = \Delta \boldsymbol{S} + \tilde{\sigma}\, (\nabla\times \boldsymbol{S})\times {\boldsymbol{S}} + \alpha\, (1 -\boldsymbol{S}^2)\boldsymbol{S} \ 
\label{eq:P_rot}
\end{equation}
The first two terms directly derive from Eq.(\ref{eq:thetas_rot}) (cf. SM \cite{SM}).
Relaxing the spherical constraint $|\boldsymbol{S}|=1$, the last Landau-like term enforces {a preferential magnitude} with a Lagrange multiplier $\alpha$. {Physically, $\alpha$ sets the defect core size $\varepsilon$ : $\alpha \sim \varepsilon^{- 2}$ .} 
Since $(\nabla \times \boldsymbol{S})\times \boldsymbol{S} = (\boldsymbol{S}\cdot \nabla ) \boldsymbol{S} - \frac{1}{2}\nabla |\boldsymbol{S}|^2$, the rotational term binds together the $\propto(\boldsymbol{S}\cdot \nabla ) \boldsymbol{S}$ self-advecting term and $\propto\nabla\,|\boldsymbol{S}|^2$ self-anchoring term of the constant density Toner-Tu equation \cite{toner1995long, toner2005hydrodynamics, deluca2024supramolecular}.  Thus, our agent-based model provides a natural microscopic origin to the phenomenological coefficients in such field theories of flocking. 
In \cite{vafa2022defect},  the same twisting of isolated positive defects was found by just considering the self-advection term in the $O(2)$ model, showing the robustness of this phenomenon. 
By setting $\tilde \sigma = 2a\sigma$ ($a$ is the lattice spacing), the numerical integration of Eq.(\ref{eq:P_rot}) provides the same annihilation times as the lattice model once rescaled by the $2d$XY value{, cf. black dots in Fig.~\ref{fig:3}(e). }
 The continuum model also faithfully reproduces the asymmetry between the $\pm 1$ defects' kinetics during annihilation, cf. Fig.~\ref{fig:3}(f,g).



\begin{figure}[b]
    \centering
    \includegraphics[width=1\linewidth]{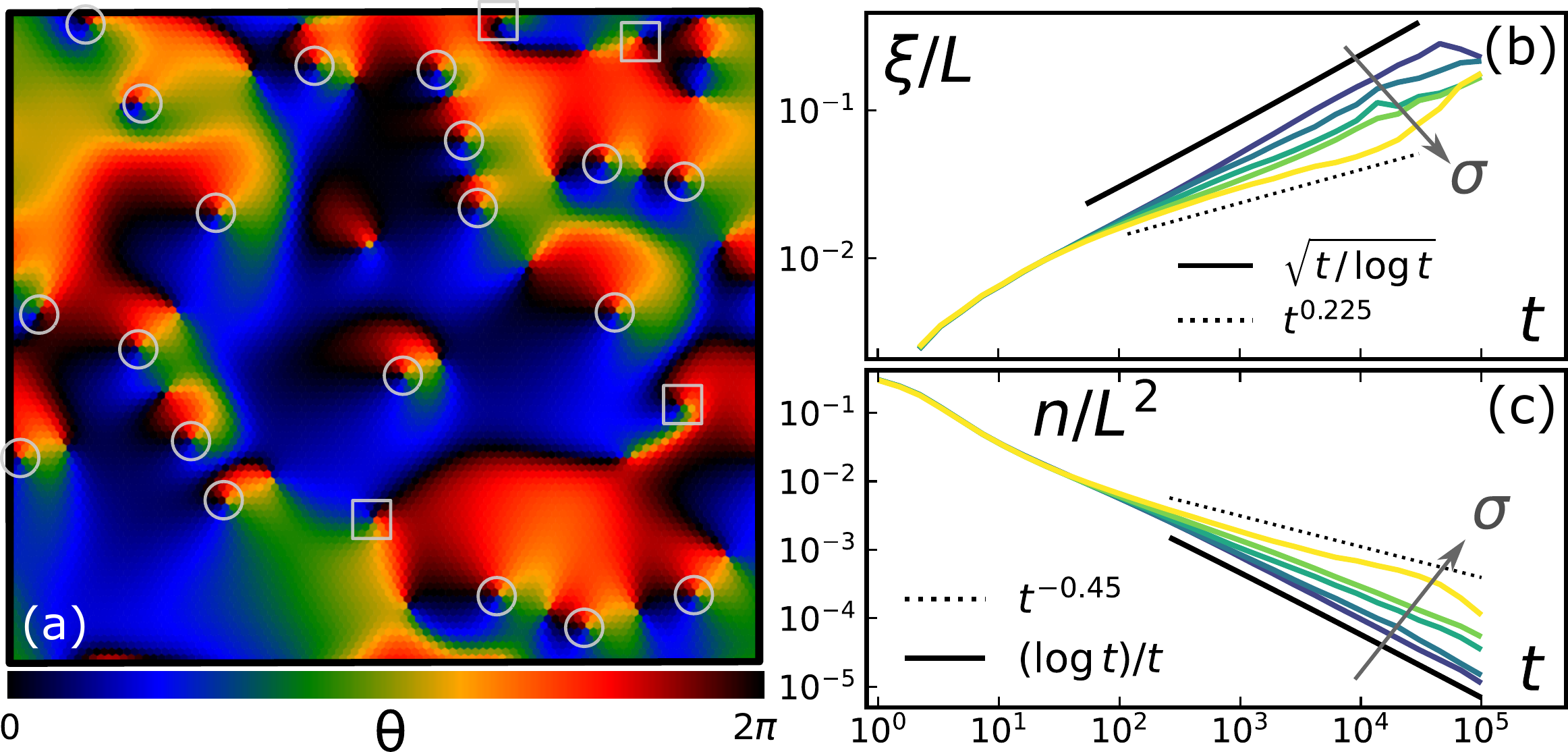}   
    \caption{ 
   \textbf{(a)} Snapshot of a $L=100$ system at $t=200, T=0, \sigma = 0.3$. 
   Sink defects ($q=+1, \mu_+=\pi$) are circled. The other $q=+1$ defects are squared. {Negative defects are not highlighted so the underlying snapshot remains visible, but the total topological charge is 0 due to PBC.  
   \textbf{(b)} Characteristic lengthscale $\xi/L$    
 and \textbf{(c)}  defect density $n/L^2$ for $L=200, T=0.08 \,T_{KT}$ and $\sigma = 0, 0.1, 0.2, 0.3, 0.4$. Black lines are the expected scalings for the equilibrium XY model. The dotted lines are powerlaw fits of the $\sigma =  0.4$ yellow curve.}  }
    \label{fig:4}
\end{figure}

As NR interactions impact significantly the defect annihilation, it is worth exploring its consequences in the coarsening process. We thus analyse the relaxation following a quench from a disordered state, with a large number of defects, to very low temperatures, where only a few bonded defects might persist \cite{rojas1999dynamical, JelicCugliandolo, rouzairedynamics}. In Fig.\ref{fig:4}(a), we show a snapshot  of the system during its relaxation at $T=0$.  
{
We define $\xi$ from the spatial correlation function $C(r)= \langle \boldsymbol{S}_i \cdot \boldsymbol{S}_j \rangle 
    = \langle \cos (\theta_i - \theta_j) \rangle $ by $C(\xi)=1/e$ and plot its time evolution over the coarsening for different values of $\sigma$ at $T=0.08\,T_{KT}$ in Fig.\ref{fig:4}(b). At equilibrium ($\sigma=0$, in blue), $\xi \sim (t/\log t)^{1/2}$, recovering the dynamical exponent $z = 2$ up to logarithmic corrections, as expected for the  $2d$XY model \cite{YurkeHuse1993, JelicCugliandolo, rouzairedynamics} (recall that $z$ is defined as $\xi\sim t^{1/z}$) . Increasing $\sigma$, the coarsening remarkably slows down and the dynamic exponent also increases up to values above $z=4$, cf. grey dotted lines. This is directly related to the slower decay of the defect density $n/L^2$ , cf. Fig.\ref{fig:4}(c). For $\sigma>0$, all the defects eventually decay to the stable sink shape, giving rise to long lived \emph{chains of} \emph{sinks} (circles in Fig.\ref{fig:4}(a)) separated by $-1$ defects with symmetry axes that hinder the annihilation with the nearest $+1$ defects. In the $2d$XY model, those structures are highly unstable and thus never observed. Non-reciprocity stabilises those large-scale structures (cf.  Movies \cite{SM}), explaining the slowing down of the decay of the defect density. 
} Occasionally, defect pairs spontaneously generate to release local excessive {stress}, but the newly created defects {always} rapidly annihilate (cf. \cite{SM} Movie 6b). As such, the steady state is typically defectless at such low temperatures, in agreement with the long-ranged ordered phase reported in a similar lattice model \cite{loos2023long}. We did not find evidence for the metastability of the ordered state reported in \cite{besse2022metastability}, that might be due to either differences between those two approaches or finite-size effects.  
\newline

To conclude, we have shown how NR interactions in the vision-cone spirit impact topological defects in a paradigmatic continuous spin system. Non-reciprocity breaks the parity and rotational symmetry of the $2d$XY model and drives the system out of equilibrium. For slightly anisotropic vision cones, non-reciprocity translates into a novel term in the continuum dynamical equation of the spin field, proportional to its vorticity. {Notably, this new term gives (and constrains) two of the three active terms in the constant density Toner-Tu equations \cite{toner1995long}, a field theory developed for motile agents. As such, our approach conceptually relates non-reciprocal systems of immobile agents and active flocks.  In the same spirit, a recent work \cite{huang2024active} also derives a hydrodynamic theory that encompasses Active Model B+ \cite{tjhung2018cluster},  
 from a microscopic model of non-motile particles interacting in a non-reciprocal fashion. 
The impact of such rotational term is to twist the phase field and to discriminate defects according to their specific shape (on top of their charge), leading to a rich phenomenology for pair annihilation: the motion of $\pm1$ defects is asymmetric, they can be speeded up or slowed down without altering the shape of the stable configuration. This is in contrast to the dynamics of liquid crystals, ruled by the relaxation of a functional, for which the relative values of bend and splay rigidity play a crucial role in the dynamics of topological defects, but cannot directly tune the relative velocity of $\pm 1$ defects without altering the stability of different shapes \cite{stannarius2016defect, missaoui2020annihilation, harth2020topological, huang2023structures}. Our intricate scenario goes well beyond the isotropic annihilation of defects in the $2d$XY model and appears to be robust and general, 
 as it remains valid at the hydrodynamic level. It neither depends  on the specific shape of the interaction kernel, as we have found an equivalence relation between sharp and soft vision cones based on its  anisotropy. 

\paragraph{Acknowledgements}
D.L. and I.P. acknowledge DURSI for financial support under Project No. 2021SGR-673. 
I.P.  and Y.R. acknowledge support from Ministerio de Ciencia, Innovaci\'on y
Universidades MCIU/AEI/FEDER for financial support under
grant agreement PID2021-126570NB-100 AEI/FEDER-EU. D.L. acknowledges MCIU/AEI for financial support under
grant agreement PID2022-140407NB-C22. 
I.P. acknowledges Generalitat de Catalunya for financial support under Program Icrea Acad\`emia.
D.J.G.P. acknowledges Swiss National Science Foundation for financial support under SNSF Starting Grant TMSGI2\_211367. 
Y.R. thanks Elisabeth Agoritsas, Boris Bergsma and Filippo de Luca for fruitful discussions. 
Y.R., D.L. and I.P thank Hugues Chat\'e and Alexandre Solon for pointing out relevant references and triggering fruitful discussions.

\bibliography{biblio}

\newpage

\appendix
\onecolumngrid
\tableofcontents

\section{Supplemental Movies}
We provide short movies to visualize the dynamics at play at the single defect level, at the pair level and finally at the system's level (unidirectional gradient propagation as reported in \cite{loos2023long} and coarsening dynamics). \\
All videos can be found on \href{https://www.youtube.com/playlist?list=PLDLpeAHw9Khf5NMAXeNH-mesFoR8pAe1B}{this Youtube channel}.
\\

It helps sketching the configuration to understand which spins are mainly influenced by which spins. Here is a correspondence that might help :\\
Black : $\theta = 0 \ (\rightarrow)$ \\
Blue : $\theta = \pi/2 \ (\uparrow)$ \\
Green/yellow : $\theta = \pi \ (\leftarrow)$ \\
Red/orange : $\theta = 3\pi/2 = -\pi/2 \ (\downarrow)$ \\

\paragraph{Single defects } \  \\
Both movies are at $T=0, \sigma = 0.3$. 
\begin{itemize}
    \item ``1- Twist" : \\ movie of the twist of a positive defect with an initial shape $\mu_+(t=0) = 1$. The final state is a sink, easily recognised by the black colour being on the left of the defect. 
    \item ``2-Polarisation" : \\ movie of the twist of a negative defect with an initial shape $\mu_-(t=0) = 1$.  
\end{itemize}

\paragraph{Pairs of defects}  \  \\
For these videos, the initial separating distance $R_0 = 48$, {the} linear size of the system $L=100$ lattice spacings. $T=0$. Hereafter, {the shapes indicated in the titles are the initial shapes at $t=0$}. 
\begin{itemize}
    \item ``3-XY annihilation" : \\Annihilation of a pair. XY equilibrium case : $\sigma=0$. Note how the $\theta$ field conserves its symmetry with respect to the defects cores {over} the process.
    \item ``4a-enhanced annihilation mu0" : \\Enhanced annihilation of a pair $\mu_+ = 0, \mu_- = \pi$ for $\sigma=0.3$ . Note how the $\theta$ field looses its equilibrium symmetry {over} the process. The polarisation axis progressively grows in the direction of the positive defects. Until now, none of the defects move in space. Then the axis reaches the $+1$ defect. At from this point, both defects run along the axis that now connects them until annihilation. 
    \item ``4b-enhanced annihilation mu0.2" : \\Enhanced annihilation of a pair $\mu_+ = 0.2, \mu_- = \pi-0.2$ for $\sigma=0.3$ . Here the initial shape of the positive defect $\mu_+ = 0.2\neq 0$ so the positive defect should, if it were isolated, decay to the sink state. It does indeed start to twist in the first instants of the movie but the negative defect rapidly polarises its surroundings: the polarisation axis reaches the positive defect before it has time to complete its twist. The two defects are now connected by the axis and eventually annihilate following the scenario described above for a pure source $\mu_+ = 0$.  
    \item     ``5a-hindered annihilation mu pi" : \\ Hindered annihilation of a pair $\mu_+ = \pi, \mu_- = 0$ for $\sigma=0.3$ . Typical configuration where the axis of the negative defect grows perpendicularly, hindering the annihilation process.
    \item ``5b-hindered annihilation mu pi-0.5" : \\ Hindered annihilation of a pair $\mu_+ = \pi-0.5, \mu_- = 2\pi - 0.5$ for $\sigma=0.3$. Since $\mu_- \neq 0,\pi$ the polarisation axis (initially)  grows obliquely. Yet, complex dynamics of the surrounding $\theta$ field impose an effective force on the axis, such that it becomes perpendicular to the core-to-core direction. Thus, it is hard for the $-1$ defect to move: the XY annihilation process is hindered.
    \item ``6a-transverse motion mu pi2" : \\ Transverse motion, for  $\mu_+ = \pi/2, \mu_- = 3\pi/2$ and $\sigma=0.3$ . The polarisation axis grows obliquely while the positive defect decays to a sink. The negative defect then follows the polarised path just created. It eventually places itself below the positive defect, because in this configuration the axis is now perpendicular to the core-to-core direction. We are now in a configuration that leads to the hindering of the annihilation process. 
    \item ``6b-transverse motion creation defects mu pi2" : \\ Same initial configuration, but now for  $\sigma=0.5$ . Note how the polarisation axis, separating two domains of opposite orientation becomes so thin that the natural elasticity of the $\theta$ field cannot support the large gradients $|\nabla \theta|$ at this location. It therefore breaks to relieve the local tension, creating a new pair of defects: the new $+1$ is on the right and meets/{annihilates} with the old $-1$. The new $-1$ follows the rest of the polarised axis and places itself below the $+1$, for the same reasons as explained above.
    \item     ``6c-transverse motion annihilation mu pi2" : \\ Same initial configuration, but now for  $\sigma=0.2, T=0.12\,T_{KT}$ . The negative defect first follows the path traced by the axis, then stabilises below the positive defect for the same reasons as explained above. The difference with the two previous videos is that the present one is at finite temperature, such that the thermal fluctuations help the defects to annihilate.
\end{itemize}
 
\paragraph{Coarsening dynamics of the system}  \  \\
We provide 3 movies of a $L=200$ triangular lattice, at $T/T_{KT} =0.08$. 
\begin{itemize}
    \item ``7a-XY system sigma 0" : \\ XY case: $\sigma =0$ The field around defects usually is isotropic and non-twisted. Defect annihilate {pairwise} following the scenario shown for isolated pairs of defects.
    \item ``7b-NRXY system sigma 0.15" : \\ Non-reciprocal case: $\sigma =0.15$ . Note how positive defects twist and how negative defects {grow} their polarisation axes, creating straight lines in the system.  The behaviours described at the level of a single pair of defect are still valid in this many-body system of defects.
    \item ``7c-NRXY system sigma 0.3" : \\Non-reciprocal case: $\sigma =0.3$ . Same {phenomenology}, but the effects are intensified.
\end{itemize}

\paragraph{Miscellaneous}  \  \\
\begin{itemize}
    \item ``8a-NRXY propagation sigma 0.5" : \\ Unidirectional propagation of gradients, as already observed by \cite{loos2023long} .
    \item  ``9-XY-NRXY chain of sinks" : \\ A chain of sinks $+1$ defects (and their corresponding $-1$ defects with $\mu_-=\pi$ ) is unstable in the equilibrium XY model (left) and stable for the NR XY model ($\sigma = 0.3$). The video is at $T=0$ for visualisation purposes but the phenomenon exists at finite $0<T<T_{KT}$. 
\end{itemize}

\section{Implementation details}
The code is written in Julia language and is accessible publicly on GitHub at \href{https://github.com/yrouzaire/NRXY-Model}{github.com/yrouzaire/NRXY-Model}.
Visit \href{https://julialang.org/}{julialang.org} to install Julia. 

\subsection{Equation of motion : adimensionalisation procedure}
Under its most general {overdamped} form, the phase of a spin $\theta_i$ follows the Langevin equation
$$
\gamma \dot{\theta}_i(t)=J \sum_{j \in \partial_i}g(\varphi_{ij})\  \sin \left(\theta_j-\theta_i\right)+\sqrt{2 k_B T \gamma}\ \eta_i(t)
$$
where $J$ is the elastic constant, $\gamma$ the damping coefficient, $k_{B}$ the Boltzmann constant and $\eta$ a unit {Gaussian} delta correlated noise. Dividing by $J$ leads to
$$
\frac{\gamma}{J} \dot{\theta}_i(t)=\sum_{j \in \partial_i}g(\varphi_{ij})\  \sin \left(\theta_j-\theta_i\right)+\sqrt{\frac{2 k_B T \gamma}{J^2}}\  \eta_i(t) .
$$
We then normalise time $\tilde{t}=t J / \gamma$ and define $\tilde{T}=k_B T / J$. One thus obtains
$$
\dot{\theta}_i(\tilde{t})=\sum_{j \in \partial_i} g(\varphi_{ij})\ \sin \left(\theta_j-\theta_i\right)+\sqrt{\frac{2 \tilde{T} \gamma}{J}} \ \eta_i(\tilde{t} \gamma / J)
$$
Finally, since the delta function in the variance of the white noise $\eta_i$ satisfies $\delta(a x)=\frac{\delta(x)}{|a|}$, one obtains the dimensionless equation of motion:
$$
\dot{\theta}_i(\tilde{t})=\sum_{j \in \partial_i} g(\varphi_{ij})\ \sin \left(\theta_j-\theta_i\right)+\sqrt{2 \tilde{T}} \ \eta_i(\tilde{t}) .
$$

We numerically integrate this equation in time using the Euler-Maruyama update rule: 
$$
\theta_i(t+dt)=\theta_i(t) + dt \sum_{j \in \partial_i} g(\varphi_{ij})\ \sin \left(\theta_j-\theta_i\right)+\sqrt{2 dt \cdot\tilde{T}} \ \eta_i 
$$
where the timestep $dt = 0.1/z$, where $z$ is the coordination number (the number of nearest neighbours) and $\eta_i \sim \mathcal{N}(0,1)$.

\subsection{Numerical Implementation of the Square and Triangular Lattices}

All the values of $\theta_i$ are stored in a square matrix for numerical efficiency, even for the triangular lattice. The topology of the network is encoded in the definition of the nearest neighbours, as shown in Fig.\ref{fig:matrix_triangular}. The indices $i=1, ..., L$ correspond to the rows and run from top to bottom. The indices $j=1, ..., L$ correspond to the columns and run from left to right. 
\begin{figure}[h!]
    \centering
    \includegraphics[width=0.6\linewidth]{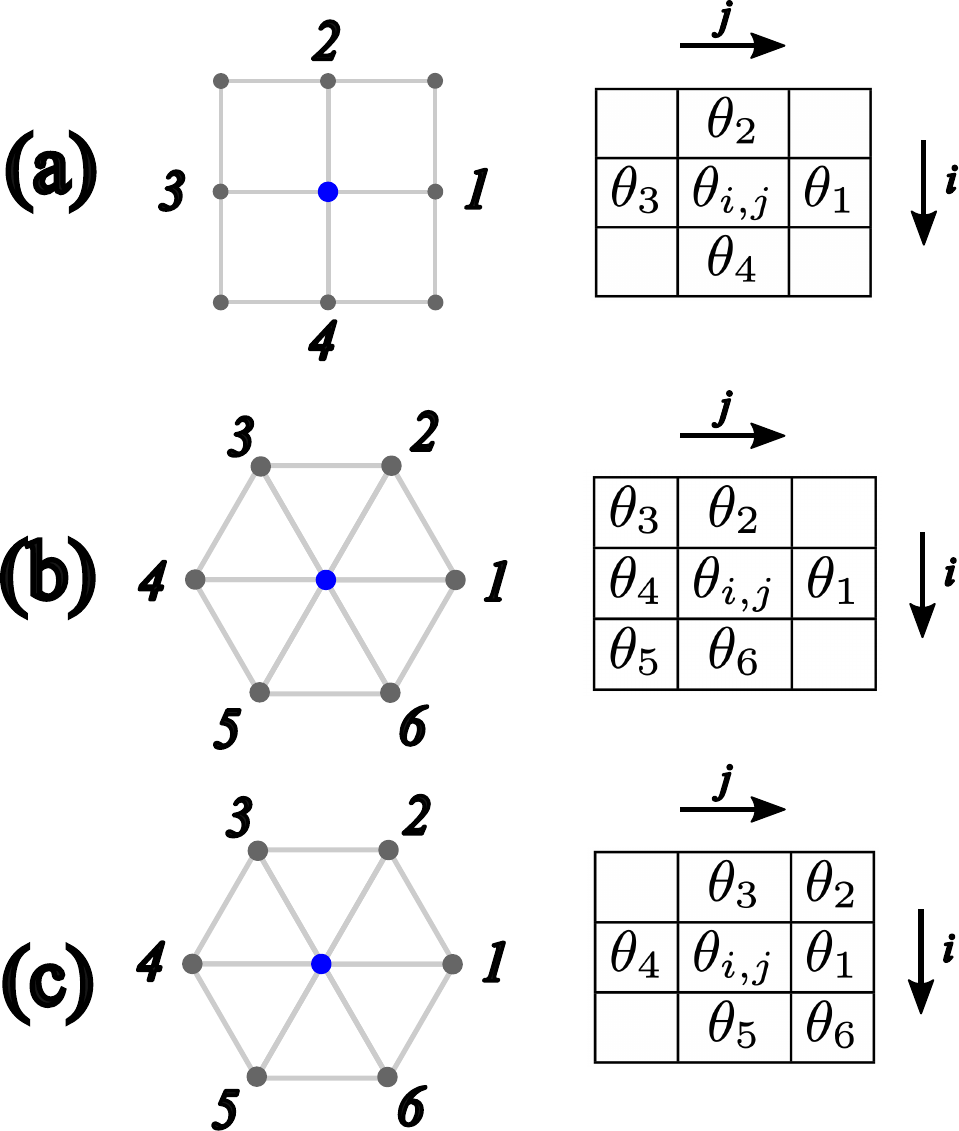}
    \caption{Nearest neighbours structure for \\
    \textbf{(a)} square lattice, for all $i,j$.\\
    \textbf{(b)} triangular lattice, for $i$ even and all $j$.\\
    \textbf{(c)} triangular lattice, for $i$ odd and all $j$.\\
    A triangular lattice can be thought as a square lattice with all odd rows laterally displaced by one half of the lattice spacing.}
    \label{fig:matrix_triangular}
\end{figure}

\subsection{Computation of the spatial correlation function}
We define the spatial correlation function
\begin{equation}
    C(r) = \langle \boldsymbol{S}_i \cdot \boldsymbol{S}_j \rangle 
    = \langle \cos (\theta_i - \theta_j) \rangle 
\end{equation}
where spins $i$ and $j$ are separated by a Euclidian distance $r$. The brackets $\langle \cdot\rangle$ denote an ensemble average {over realisations of the thermal noise and over all spins}. Computing directly such a quantity in a $L\times L$ system scales as $\mathcal{O}(L^3)$, which is prohibitive for large systems. For numerical efficiency, we compute it in the Fourier space, which reduces the numerical cost to $\mathcal{O}(L^2\,\log L)$. 
First note that, by trigonometric identities, 
\begin{equation}
    \cos (\theta_i - \theta_j) = \cos (\theta_i)\cos (\theta_j) + \sin (\theta_i)\sin(\theta_j)
\end{equation}
The term $\cos (\theta_i)\cos (\theta_j)$ is a convolution and can be easily handled in Fourier space. 

We note the Fourier transform operator $\mathcal{F}$ and  its inverse $\mathcal{F}^{-1}$. 
We note ``$.^2$" the element-wise square operator: if $x$ is a vector of components $x_i$,  then $x.^2$ is a vector of the same length of components $(x_i)^2$  . 

With this notation in mind, the $2d$ spatial correlation function can be expressed as 
\begin{equation}
    C_{2d} = C(r_x, r_y) = | \mathcal{F}^{-1}(|(\mathcal{F}[\cos \theta]) .^2 + (\mathcal{F}[\sin \theta]) .^2|  ) |
\end{equation}
In a continuum simulation, $ C_{2d} $ would be a radial function. Here, because of the discrete nature of the underlying lattice, $ C_{2d} $ is slightly anisotropic. To obtain the usual unidimensional $C(r)$, we take the mean of the two principal components (along the axes) and normalise to get $C(r=0)=1$.

\subsection{Constructing a pair of defects with controlled shapes}
The first step to investigate an isolated pair of defects is to be able to create 2 defects with controlled shapes $\mu_+$ and $\mu_-$.
\begin{figure}[h!]
    \centering
    \includegraphics[width=0.5\linewidth]{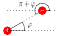}   
    \caption{ 
Sketch of a pair of defects and definition of the polar angle $\varphi$ that the negative defect makes with respect to the positive defect. 
    }
    \label{fig:creation_pair}
\end{figure}
To do so, recall that around an ideal topological defect, $\theta = q\arctan(y/x) + \mu$. In polar coordinates, where $\varphi=\arctan(y/x)$ (depicted in Fig.\ref{fig:creation_pair}), the field contribution from the defect reads $\theta = q\varphi + \mu$. 

It follows that if one creates a positive defect with $\mu_1$ and a negative defect with $\mu_2$, both fields will add up. 
At the location of the negative defect, the sum of the two contributions is 
\begin{equation}
    \mu_- = \left[\mu_1 + (+1)\varphi \right] + \left[\mu_2\right]  = \mu_1 + \mu_2 + \varphi \ .
    \label{eq_app:mu-}
\end{equation}
At the location of the positive defect, the sum of the two contributions is 
\begin{equation}
    \mu_+ =  \left[\mu_1 \right] + \left[(-1)(\varphi+\pi) +\mu_2\right] = \mu_1 + \mu_2 - \pi -\varphi \ .
    \label{eq:mu-}
\end{equation}
Those are the concrete equations we use to create defects with controlled $\mu_+,\mu_-,\varphi$ (only 2 of these 3 parameters are independent), tuning $\mu_1 + \mu_2$ to obtain the desired $\mu_+,\mu_-$.  \\
Also, one can get rid of the artificial $\mu_1,\mu_2$ by combining these equations to obtain: 
\begin{equation}
    \mu_- - \mu_+ =  \pi + 2\varphi \ .
    \label{eq_app:geom_adequation}
\end{equation}
{Since the shapes are equivalent modulo $2\pi$, Eq.(\ref{eq_app:geom_adequation}) can also be written }
\begin{equation}
    \mu_+ - \mu_- =  \pi - 2\varphi \ .
    \label{eq_app:geom_adequation2}
\end{equation}
As it creates the smoothest possible field, this equation represents the state of least distortion. It also highlights that only two of these three parameters are independent.
In the main text, and without loss of generality, we work with $\varphi=0$: the defects are horizontally aligned and the positive defect is on the left side. 

\section{The defects' shape : definition, comments and analytic calculations}
The field around a topological charge is given by the solution of the $2d$ Poisson equation: $\theta(x,y) = q\,\arctan(y/x) + \mu$. \ylann{Due to the polar symmetry of the model (as opposed to nematic), only integer defects are allowed. Defects with a topological charge $|q|\geq 2$ are not stable and very rapidly decay to $|q|=1$ defects, such that we will only consider those in what follows. Once a $q=\pm1$} defect is located, i.e. one knows $(x_{core},y_{core},q)$; we then naturally define $\mu_\pm = \text{Arg}\left\{\sum_j \exp(i(\theta_j - q\,u_j))\right\}$, where the sum runs over the spins surrounding the plaquette of the dual lattice where the defect lies, $i$ is the imaginary unit and Arg is the function that returns the phase of a complex number: Arg$(re^{i\theta}) = \theta$. 
Each neighbour $j$ in that sum makes a certain angle $u_j= \arctan(y_j/x_j)$ with respect to the horizontal axis. Irrespective of its position, we note its orientation $\theta_j$. We sketch this in Fig.\ref{fig:plaquettes}, where the red dots are the defects cores. For example, for the square lattice (left panel), $u_j = j\pi/4, \ j=1,2,3,4$. For the upward (inner) triangular plaquette (central panel),  $u_{1,2,3} = \pi/2,-5\pi/6,-\pi/6$.
If the defect is located on a downward triangular plaquette (right panel), one has $u_{1,2,3} = 5\pi/6,-\pi/2,\pi/6$. 

\begin{figure}[h!]
    \centering
    \includegraphics[width=0.8\linewidth]{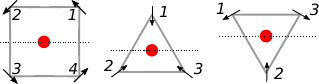}
    \caption{Defects (red) lie at the centre of the dual plaquettes while the spins live on the lattice nodes (grey). Left panel: the unique plaquette of the square lattice. Centre and right panels: the two plaquettes (resp. up and down) of the triangular lattice. Dotted lines are horizontal and pass through the defects; the angles $\theta_j$ and $u_j$ are expressed with respect to that axis in the counter-clockwise direction. }
    \label{fig:plaquettes}
\end{figure}

Note that $\mu$ can be thought of as the value of the $\theta$-field at the immediate right side of the defect, as sketched in Fig.\ref{fig:square_plaquette_spin_right} with the imaginary green arrows.
\begin{figure}[h!]
    \centering
    \includegraphics[width=0.85\linewidth]{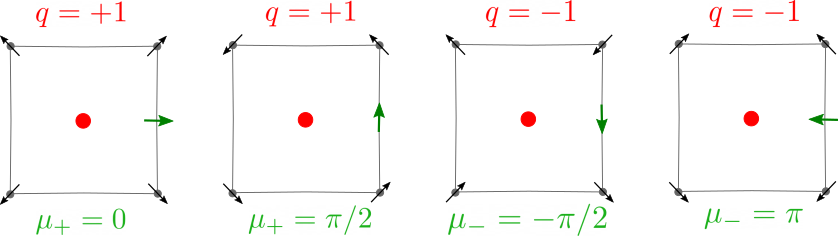}
    \caption{The defects (red) live at the center of the dual plaquettes. The spins (black) live on the lattice, i.e. on the corners of the plaquette. An easy way to visually identify $\mu$ is to imagine the orientation an imaginary spin (green) would have if  located at the right edge of the defect ($y=0$ and thus $\arctan (y/x) = 0$).}
    \label{fig:square_plaquette_spin_right}
\end{figure}

To support the intuitive argument sketched in Fig.2(e, l) of the main text, we analytically compute the dynamics of the field around ideal isolated topological defects.
We start from the  equation of motion
	\begin{equation}
		\gamma\,\dot{\theta}_i=J\sum_{j \in \partial_i} \,g_{\sigma}(\varphi_{i j})\sin \left(\theta_j-\theta_i\right) + \sqrt{2\gamma\, T}\,\eta_i(t) 
	\end{equation}

	We first study this dynamics in the continuum limit, where the orientation field is a function of a position $\underline{x}$, and consider a neighbourhood around this position. We also ignore noise and set $J/\gamma = 1$ for simplicity.
	\begin{equation}
		\dot{\theta}(\underline{x}) = \int_{-\pi}^{\pi}g_{\sigma}(\phi_1 - \theta(\underline{x}))\sin \left(\theta(\underline{x} + \delta \hat{\underline{\phi}}_1)-\theta(\underline{x})\right) d\phi_1
	\end{equation}
	
	Where $\phi_1$ is the polar angle around position $\underline{x}$ and $\delta$ is a small finite distance which could be thought of as the interdefect spacing and $\hat{\underline{\phi}}_1$ is a unit vector with orientation $\phi_1$. {Approximating $\theta(\underline{x} + \delta \hat{\underline{\phi}}_1)$ by its first order Taylor expansion, one gets}:
	
	\begin{align}
		\dot{\theta}(\underline{x}) &= \int_{-\pi}^{\pi}g_{\sigma}(\phi_1 - \theta(\underline{x}))\sin \left(\theta(\underline{x}) + \delta \hat{\underline{\phi}}_1\nabla\theta(\underline{x})-\theta(\underline{x})\right) d\phi_1\\
		&= \int_{-\pi}^{\pi}g_{\sigma}(\phi_1 - \theta(\underline{x}))\sin \left(\delta \hat{\underline{\phi}}_1\nabla\theta(\underline{x})\right) d\phi_1		
	\end{align}

	We now convert our coordinate system to polar coordinates around the center of a defect, which we denote $(\phi,r)$ with corresponding orthonormal basis vectors $[\underline{\hat{r}},\underline{\hat{\phi}}]$ shown in the figure below. 
	\begin{figure}[h!]
		\centering
		\includegraphics[width=0.35\columnwidth]{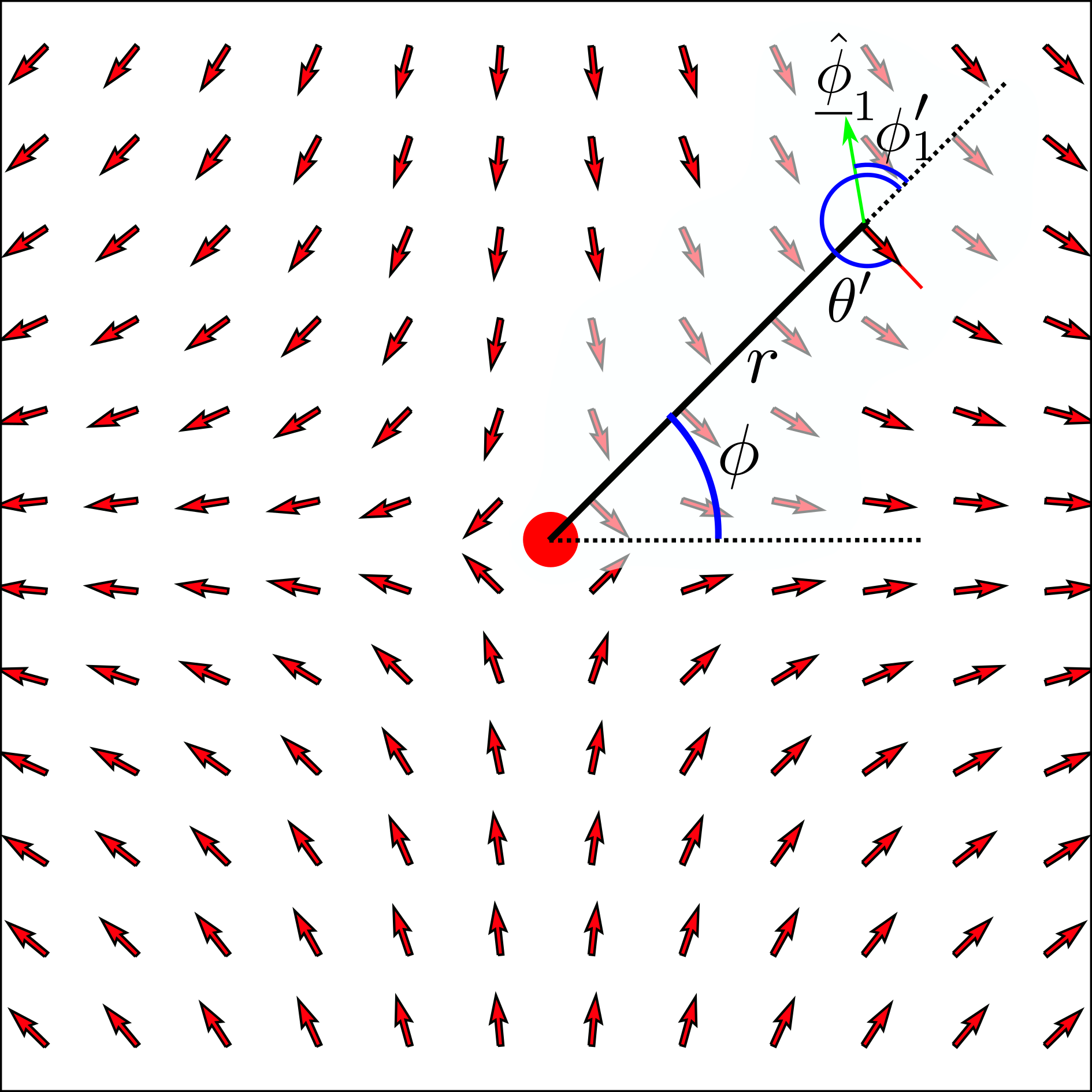}\caption{Coordinate system for the calculation.\label{fig:Coords}}
	\end{figure}

	In this coordinate system, the new director field can now be written relative to the $\underline{\hat{r}}$ direction as:
	\begin{equation}
		\theta'(r,\phi) = \theta(\underline{x}) - \phi \implies \dot{\theta}' = \dot{\theta}
	\end{equation}
	 We must also convert the coordinate system over which we are performing the integration, thus:
	 \begin{align}
	 	\phi_1' &= \phi_1 - \phi\\
	 	\hat{\underline{\phi}}_1 &= \cos(\phi_1')\underline{\hat{r}} +  \sin(\phi_1')\underline{\hat{\phi}}
	 \end{align}
	 
	This leaves us at:
	\begin{equation}
		\dot{\theta}' = \int_{-\pi}^{\pi}g_{\sigma}(\phi_1' - \theta')\sin \left(\delta \hat{\underline{\phi}}_1\nabla\theta(\underline{x})\right) d\phi_1'
	\end{equation}
	Note, we have not changed the limits of the integration as all functions are periodic over the interval $[-\pi,\pi]$ so shifting by a constant does not change the integral.

	Finally, we assume that $\theta$ does not vary radially around a topological defect, thus $\partial_r\theta' = 0$. This is true for an isolated defect in the reciprocal XY model and allows us to write the gradient of $\theta$ in the polar basis:	 
	\begin{equation}
 		\nabla\theta = \frac{(\partial_\phi\theta' + 1)}{r}\underline{\hat{\phi}}
	\end{equation}

	Which gives us
	\begin{equation}
		\nabla\theta.\underline{\hat{\phi}}_1 = \frac{(\partial_\phi\theta' + 1)}{r}\sin(\phi_1)
	\end{equation}

	Using these identities we can convert our equation to the new coordinate system to arrive at:
	\begin{equation}
		\dot{\theta} = \int_{-\pi}^{\pi}g_{\sigma}(\phi_1 - \theta)\sin \left(\delta (\partial_\phi\theta + 1)\frac{\sin(\phi_1)}{r} \right) d\phi_1		
		\label{eq:E}
	\end{equation}
	Where we have dropped all $'$. 
	
	We consider the non-reciprocal kernel as $g_{\sigma}(\varphi) = \exp(\sigma\cos(\varphi)) \approx 1 + \sigma\cos(\varphi)$ for small $\sigma$. 
	\begin{align}
		\dot{\theta} &= \int_{-\pi}^{\pi}[1+\sigma\cos(\phi_1 - \theta)]\sin \left(\delta (\partial_\phi\theta + 1)\frac{\sin(\phi_1)}{r} \right) d\phi_1	\\
		&= \int_{-\pi}^{\pi}\sin \left(\delta (\partial_\phi\theta + 1)\frac{\sin(\phi_1)}{r} \right) d\phi_1 + \sigma\int_{-\pi}^{\pi}\cos(\phi_1 - \theta)\sin \left(\delta (\partial_\phi\theta + 1)\frac{\sin(\phi_1)}{r} \right) d\phi_1	
	\end{align}

	\subsection{Stability of a +1 defect}
	
	For a $+1$ defect, $\partial_\phi\theta' = 0$ and $\theta' = \mu$. With $\mu = 0$ signifying a source and $\mu = \pi$ being a sink. In this scenario there is no variance on the polar angle around the defect and we can write a single update equation for the shape of the defect $\mu$.
	
	\begin{equation}
	\dot{\mu} = \cancelto{0}{\int_{-\pi}^{\pi}\sin \left(\delta \frac{\sin(\phi_1)}{r} \right) d\phi_1} + \sigma\int_{-\pi}^{\pi}\cos(\phi_1 - \mu)\sin \left(\delta\frac{\sin(\phi_1)}{r} \right) d\phi_1	
	\end{equation}

	Since $\sin \left(\delta\frac{\sin(\phi_1)}{r} \right)$ is anti-symmetric about $\phi_1=0$, the second integral will only give zero in the case where $\cos(\phi_1 - \mu)$ is symmetric about zero. This is only true for $\mu = 0, \pi$, thus the only fixed points of this equation are these values. 
	
	We now linearize this equations around $\mu$ using angle addition formula.
	\begin{align}
		\dot{\Delta\mu} &= \sigma\int_{-\pi}^{\pi}\cos(\phi_1 - \mu - \Delta\mu)\sin \left(\delta\frac{\sin(\phi_1)}{r} \right) d\phi_1 - \dot{\mu}\\
		& = \sigma\int_{-\pi}^{\pi}[\cos(\phi_1-\mu)\cos(\Delta\mu) + \sin(\phi_1-\mu)\sin(\Delta\mu)]\sin \left(\delta\frac{\sin(\phi_1)}{r} \right) d\phi_1 - \dot{\mu}\\
		&= \cancel{\sigma\int_{-\pi}^{\pi}\cos(\phi_1-\mu)\sin \left(\delta\frac{\sin(\phi_1)}{r} \right) d\phi_1} + \Delta\mu\sigma\int_{-\pi}^{\pi}\sin(\phi_1-\mu)\sin \left(\delta\frac{\sin(\phi_1)}{r} \right) d\phi_1 - \cancel{\dot{\mu}}\\
		&= \Delta\mu\sigma\int_{-\pi}^{\pi}\sin(\phi_1-\mu)\sin \left(\delta\frac{\sin(\phi_1)}{r} \right) d\phi_1
	\end{align}

	This is solved trivially by the equation $\Delta\mu = A\exp[\omega t]$ with 
	\begin{equation}
		\omega = \sigma\int_{-\pi}^{\pi}\sin(\phi_1-\mu)\sin \left(\delta\frac{\sin(\phi_1)}{r} \right) d\phi_1
	\end{equation}
	Which indicates stability for $\mu = \pi$ (sink) and instability for $\mu = 0$ (source). The magnitude of $\omega$ is greatest around $r/\delta = 0.5$, which means the evolution speed of $\mu$ decreases as $r$ increases (above the lattice spacing).

	\subsection{Shape of a -1 defect}

	For a $-1$ defect, $\theta = -2\phi$ minimizes the energy in the reciprocal XY model. This is a defect in which the director points out along the $x$ axis like the one shown in Fig.~\ref{fig:Coords}. This gives 
	\begin{equation}
		\dot{\theta}(\phi) = \cancelto{0}{\int_{-\pi}^{\pi}\sin \left(-\delta \frac{\sin(\phi_1)}{r} \right) d\phi_1} + \sigma\int_{-\pi}^{\pi}\cos(\phi_1 + 2\phi)\sin \left(-\delta \frac{\sin(\phi_1)}{r} \right) d\phi_1	
	\end{equation}
	Once again the reciprocal terms vanish by symmetry. We can expand the second term using angle addition formula. 
	
	\begin{align}
		\dot{\theta}(\phi) &= \sigma\int_{-\pi}^{\pi}\cos(\phi_1 + 2\phi)\sin \left(-\delta \frac{\sin(\phi_1)}{r} \right) d\phi_1	\\
		&= \sigma\int_{-\pi}^{\pi}[\cos(\phi_1)\cos(2\phi) - \sin(\phi_1)\sin(2\phi)]\sin \left(-\delta \frac{\sin(\phi_1)}{r} \right) d\phi_1	\\
		&= \cos(2\phi)\sigma\cancelto{0}{\int_{-\pi}^{\pi}\cos(\phi_1)\sin \left(-\delta \frac{\sin(\phi_1)}{r} \right) d\phi_1} - \sin(2\phi)\sigma\int_{-\pi}^{\pi}\sin(\phi_1)\sin \left(-\delta \frac{\sin(\phi_1)}{r} \right) d\phi_1	\\
		&= \sin(-2\phi)\sigma\int_{-\pi}^{\pi}\sin(\phi_1)\sin \left(-\delta \frac{\sin(\phi_1)}{r} \right) d\phi_1
	\end{align}
	
	This has quadrupole-like symmetry such that $\theta$ will grow in the first and third quadrant, and shrink in the second and fourth, leading to a nematic symmetry of the -1 defects.
	It is not easy to perform a similar linear stability analysis for the $-1$ defect as the solution for $\dot{\theta} = 0$ is non-trivial.
	
	\subsection{Numerical results}
	
	We can estimate the stable solutions of Eq.~\ref{eq:E} for different winding numbers using finite different methods. The winding number is fixed by the initial conditions and $\theta$ is evolved using Eq.~\ref{eq:E} until a stable solution is reached. The results are shown in Fig.~\ref{fig:Defs} below.

	\begin{figure}[h]
		\centering
		\includegraphics[width=0.75\columnwidth]{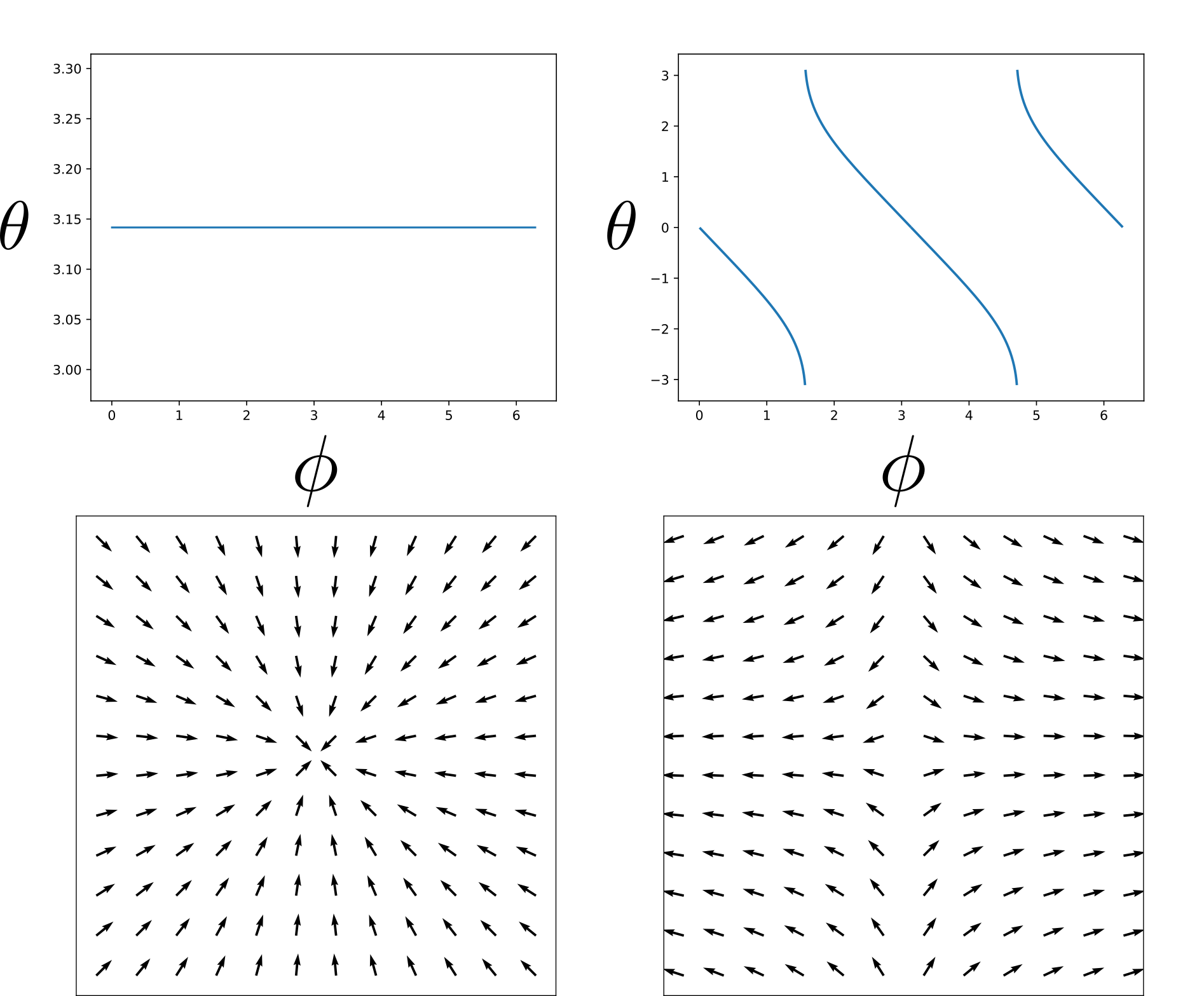}\caption{Stable configurations for +1 (left) and -1 (right) defect. Both solutions obtained by numerically solving Eq.~\ref{eq:E}. \label{fig:Defs}}
	\end{figure}

\section{Symmetries of the system}
The equilibrium XY model is invariant under global rotation of the spins $\{ \theta \} \to \{ \theta + \alpha\} $  because the alignment term between two spins $i$ and $j$ only depends on the difference of the spins orientation: 
\begin{equation}
    \sin ((\theta_j + \alpha) - (\theta_i + \alpha)) =  \sin (\theta_j - \theta_i)
\end{equation}
The reciprocal XY model is thus invariant under the addition of a global phase -- we say it is $O(2)$ invariant or, equivalently, $O(2)$ symmetric-- which prevents defects of the same topological charge but of different shapes (different phases) to behave differently.

In our 2D spin system, the parity transformation $\begin{pmatrix}
x \\ y 
\end{pmatrix}
\to 
\begin{pmatrix}
-x \\ y 
\end{pmatrix}$ can be re-expressed in terms of angles with the transformation $\{ \theta \} \to \{ -\theta + \pi\} $. Since the $2dXY$ model is $O(2)$ symmetric, we only have to check that the equation of motion is invariant under the transformation $\{ \theta \} \to \{ -\theta\} $ 
\begin{equation}
    \begin{aligned}
        \frac{d}{dt}(-\theta_i) = &\  \sin (-\theta_j - (-\theta_i)) \\ 
        -\frac{d}{dt}(\theta_i) = &\  -\sin (\theta_j - \theta_i) \\ 
        \frac{d}{dt}(\theta_i) = &\  \sin (\theta_j - \theta_i) \\
         \dot\theta_i = &\  \sin (\theta_j - \theta_i) 
    \end{aligned}
\end{equation}
So the XY model is indeed invariant under parity. Note that the parity transformation transforms a $q=+1$ defect into a $q=-1$ and vice-versa, such that XY defects behave in the same fashion regardless of their charge.

On the other hand, the non-reciprocal kernel $g$ violates all the previous symmetries. Indeed, while the alignment term depends on the difference of the spins orientation $\theta_j - \theta_i$, $g$ only depends on the orientation of the spin $i$. Therefore, $g(\theta) \neq g(\theta+\alpha)$ and the system is no longer $O(2)$ invariant. This is consistent with the results of the main text, where different shapes of the same topological charge (sinks and sources for $+1$  defects for example) behave in a totally different fashion.
In the same spirit, $g(\theta) \neq g(-\theta)$, breaking parity. This is consistent with the difference in the dynamics between $\pm 1$ defects in the non-reciprocal case. For instance, in the enhanced annihilation mechanism, positive defects move faster than negative defects, {see Fig.3(f,g) of the main text. On the other hand, in the general case (ie. for $\mu_+ \neq 0,\pi$) negative defects move faster than positive defects, see Fig.3(c) of the main text and Supplementary Movie 6a, 6b. 

Both parity and rotation invariances} are true for all kernels that do not depend on $\theta_j - \theta_i$ but only on $\theta_i$.

\section{Derivation of Eq.~(2) and Eq.~(3) of the main text}
In this section, we detail the steps to write the equation of motion $\dot{\theta}_i= \sum_j \sin \left(\theta_j-\theta_i\right)\ g(\varphi_{ij})$ in a continuous form, based on derivatives of the  field $\theta = \theta(x,y)$, which emphasizes the reciprocal and non-reciprocal contributions of the kernel. 

Consider a spin $i$ on the square lattice: it has 4 nearest neighbouring spins $j=1,2,3,4$ (right, top, left, bottom), the orientations of which we note $\theta_{1,2,3,4}$ . Those 4 neighbours can be thought to lie on a unit circle centred on the spin $i$. We call $u_{j}$ the angle formed by horizontal axis and the location 
of spin $j$, see Fig.1a of the main text. In a square lattice, $u_j = \frac{\pi}{2}(j-1), \ j=1,...,4$. In a triangular lattice, $u_j = \frac{\pi}{3}(j-1), \ j=1,...,6$ . Note that $u_j$ and the orientation $\theta_j$ are independent.

We start by explaining how \emph{not} to proceed. One could be tempted to identify, from the beginning, $\theta_j-\theta_i$ with the first order derivative of $\frac{\partial \theta}{\partial k}$ ($k=x,y$). The issue is, if one takes the simplest case of the XY model, where the kernel $g=1$, one would obtain 
\begin{equation}
    \begin{aligned}
\dot{\theta}_i= & \sum_j \sin \left(\theta_j-\theta_i\right) \\ 
= & \  \sin\left(\frac{\partial \theta}{\partial x}\right) +  \sin\left(\frac{\partial \theta}{\partial y}\right)+ \sin\left(-\frac{\partial \theta}{\partial x}\right)+  \sin\left(-\frac{\partial \theta}{\partial y}\right)\\
= & \  0 \
\end{aligned}
\end{equation}
for all fields $\theta(x,y)$ without any assumption. The reason why all terms cancel is that implicitly, one here uses the non-centred numerical definition of derivatives, therefore saying that $\frac{\partial \theta}{\partial x} = \theta_1-\theta_i = \theta_i-\theta_3$, where the last equality is incorrect in general and especially around a topological defect. 

To avoid this problem, one has to go to second order (not surprising if one wants to account for an elastic term) and consider both the spins on the right-left (resp. top-bottom) of the spin $i$ of interest.  

{In the following derivations of Eq.~(2) and Eq.~(3) of the main text, we work under the assumptions of \textit{small anisotropy/non-reciprocity }$\sigma \ll 1$ and \textit{small spatial gradients} of the field $\theta$.
The limit $\sigma \ll 1$ allows to linearise the  coupling kernel to first order in $\sigma$ ($e^{\sigma \cos x} \approx 1+\sigma \cos x$), such that :}
\begin{equation}
    \begin{aligned}
\frac{\gamma}{J} \dot{\theta}_i= & \sum_j \sin \left(\theta_j-\theta_i\right)\ g(\varphi_{ij}) \\
= & \sum_j \sin \left(\theta_j-\theta_i\right)\left[1+\sigma \cos \left(\varphi_{ij}\right)\right]\\
= & \sum_j \sin \left(\theta_j-\theta_i\right)\left[1+\sigma \cos \left(\theta_i-u_{ j}\right)\right]\\
= & \sum_j \sin \left(\theta_j-\theta_i\right)\left[1+\sigma \cos \left(\theta_i-\frac{\pi}{2}(j-1)\right)\right] \\
= & \sin \left(\theta_{1}-\theta_i\right)\left(1+\sigma \cos \theta_i\right) \\
& +\sin \left(\theta_2-\theta_j\right)\left[1+\sigma \cos \left(\theta_i-\frac{\pi}{2}\right)\right] \\
& +\sin \left(\theta_3-\theta_i\right)\left[1+\sigma \cos \left(\theta_i-\pi\right)\right] \\
& +\sin \left(\theta_4-\theta_i\right)\left[1+\sigma \cos \left(\theta_i-\frac{3 \pi}{2}\right)\right] \\
= & \sin \left(\theta_1-\theta_i\right)\left(1+\sigma \cos \theta_i\right) \\
& +\sin \left(\theta_2-\theta_i\right)\left(1+\sigma \sin \theta_i\right) \\
& +\sin \left(\theta_3-\theta_i\right)\left(1-\sigma \cos \theta_i\right) \\
& +\sin \left(\theta_4-\theta_i\right)\left(1-\sigma \sin \theta_i\right) \\
= & \sin \left(\theta_1-\theta_i\right) + \sin \left(\theta_3-\theta_i\right) + \sin \left(\theta_2-\theta_i\right) + \sin \left(\theta_4-\theta_i\right) \\
& + \sigma  \cos(\theta_i)\big(\sin \left(\theta_1-\theta_i\right)-\sin \left(\theta_3-\theta_i\right))\\
& + \sigma \sin(\theta_i)\big(\sin \left(\theta_2-\theta_i\right)-\sin \left(\theta_4-\theta_i\right))\\
\end{aligned}
\end{equation}
One has (the same goes for the vertical direction with $\theta_2$ and $\theta_4$) :
\begin{equation}
     \begin{aligned}
   &\sin \left(\theta_1-\theta_i\right) + \sin \left(\theta_3-\theta_i\right)\\
    &=2\sin \left[\frac{1}{2}\left(\theta_1-2 \theta_i+ \theta_3\right)\right] \cos \left[\frac{1}{2}\left(\theta_1-\theta_3\right)\right]
    \end{aligned}
\end{equation}
\begin{equation}
         \begin{aligned}
&\sin \left(\theta_1-\theta_1\right)-\sin \left(\theta_3-\theta_1\right)\\
&=2 \cos \left[\frac{1}{2}\left(\theta_1-2 \theta_i+\theta_3\right)\right]\sin \left[\frac{1}{2}\left(\theta_1-\theta_3\right)\right]
     \end{aligned}
\end{equation}

Now, if one identifies the differences in the continuum as
\begin{equation}
    \begin{aligned}
& a\frac{\partial \theta}{\partial x}=\frac{\theta_1-\theta_3}{2} \\
& a\frac{\partial \theta}{\partial y}=\frac{\theta_2-\theta_4}{2} \\
& a^2\frac{\partial^2 \theta}{\partial x^2}=\theta_1-2 \theta_i+\theta_3 \\
& a^2\frac{\partial^2 \theta}{\partial y^2}=\theta_2-2 \theta_i+\theta_4
\end{aligned}
\end{equation}
one obtains 
\begin{equation}
\begin{aligned}
\frac{\gamma}{J} \frac{\dot{\theta}}{2} & = \cos (a\,\theta_x) \sin (\frac{a^2}{2}\,\theta_{x x})+\cos (a\,\theta_y) \sin  (\frac{a^2}{2}\,\theta_{y y}) \\
& +\sigma\left(\cos \theta \sin (a\,\theta_x) \cos  (\frac{a^2}{2}\theta_{x x})+\sin \theta \sin (a\,\theta_y) \cos  (\frac{a^2}{2}\theta_{y y})\right)
\end{aligned}
    \label{eq:SM_theta_cos}
\end{equation}
where $\theta =\theta(x,y)$, $\theta_k=\frac{\partial \theta}{\partial k}$ and $\theta_{kk}=\frac{\partial^2 \theta}{\partial k^2}, k=x,y$.

 From this expression, it is now clear that (i) $\sigma$ cannot be absorbed in the time unit {(ii) the dynamics is $\mu$-invariant (it only depends on derivatives of the field $\theta$) if and only if $\sigma=0$}. 

{Now using the small gradients approximation, and} $\cos \alpha \approx 1$ and $\sin \alpha \approx \alpha$,  
one obtains 
\begin{equation}
    \frac{\gamma}{J} \dot{\theta} =  a^2(\theta_{x x}+\theta_{y y}) + 2a\sigma\, (\cos (\theta) \cdot \theta_x +\sin (\theta)\cdot \theta_y)
\end{equation}

The first two terms give the Laplacian operator $\Delta$ that physically represents the elasticity of the $\theta$ field.
The last two terms can be rewritten as follows: 
\begin{equation}
\cos (\theta) \cdot \theta_x +\sin (\theta)\cdot \theta_y  = \partial_x (\sin \theta) - \partial_y(\cos \theta) = (\nabla \times \boldsymbol{S})_z
\end{equation}
where $\boldsymbol{S}=(\cos\theta,\sin\theta)$ is the spin vector field, thus leading to the result of the main text Eq.~(2):
\begin{equation}
    \frac{\gamma}{J} \dot{\theta} =  a^2\,\Delta \theta + 2a\sigma\, (\nabla \times \boldsymbol{S})_z
    \label{eq:rotational}
\end{equation}
Note that here we have kept $\gamma, J$ and $a$ , even though $\gamma=J=a=1$ in the adimensionalized main text version, to highlight the unitless nature of $\sigma$. 

\ \newline 
Inspired by the formulation in Eq.~(\ref{eq:rotational}) (Eq.~2 of the main text) in terms of the field 
\begin{equation}
\begin{aligned}
\boldsymbol{S} & =\left(\begin{array}{c}
S_x \\
S_y
\end{array}\right)=\left(\begin{array}{c}
\cos \theta \\
\sin \theta
\end{array}\right)\ , 
\end{aligned}
\end{equation}
we look for an equivalent formulation entirely in terms of $\boldsymbol{S}$. Importantly, for now the magnitude of $\boldsymbol{S}$ still remains strictly fixed to $|\boldsymbol{S}|=1$, such that $\dot{\boldsymbol{S}} \perp \boldsymbol{S}$. 
Indeed, one has: 
\begin{equation}
\begin{aligned}
\dot{\boldsymbol{S}} & =\dot{\theta}\left(\begin{array}{c}
-\sin \theta \\
+\cos \theta
\end{array}\right)=\dot{\theta}\left(\begin{array}{l}
-S_y \\
+S_x
\end{array}\right)
\end{aligned}
\end{equation}

We start by the elasticity term $\Delta \boldsymbol{S}=\left(\begin{array}{c}
\Delta S_x \\
\Delta S_y
\end{array}\right)$: 
\begin{equation}
    \begin{aligned}
& \frac{\partial^2}{\partial x^2} S_x=\frac{\partial^2}{\partial x^2}(\cos \theta) \\
& =\frac{\partial}{\partial x}\left(-\frac{\partial \theta}{\partial x} \sin \theta\right) \\
& =-\left(\frac{\partial^2 \theta}{\partial x^2} \sin \theta+\left(\frac{\partial \theta}{\partial x}\right)^2 \cos \theta\right) \\
& =-\left(\frac{\partial^2 \theta}{\partial x^2}S_y+\left(\frac{\partial \theta}{\partial x}\right)^2 S_x\right) \\
& =-\frac{\partial^2 \theta}{\partial x^2} S_y \\
&
\end{aligned}
\end{equation}
Where the last equality results from the approximation that we only consider first order gradients.
Also,
\begin{equation}
    \begin{aligned}
\frac{\partial^2}{\partial x^2} S_y & =\frac{\partial^2}{\partial x^2}(\sin \theta) \\
& =\frac{\partial}{\partial x}\left(\frac{\partial \theta}{\partial x} \cos \theta\right) \\
& =\frac{\partial^2 \theta}{\partial x^2} \cos \theta-\left(\frac{\partial \theta}{\partial x}\right)^2 \sin \theta \\
& =\frac{\partial^2 \theta}{\partial x^2} S_x-\left(\frac{\partial \theta}{\partial x}\right)^2 S_y \\
& = \frac{\partial^2 \theta}{\partial x^2} S_x
\end{aligned}
\end{equation}

Similarly, $\frac{\partial^2}{\partial y^2} S_x =-\frac{\partial^2 \theta}{\partial y^2} S_y $ and $\frac{\partial^2}{\partial y^2} S_y =+\frac{\partial^2 \theta}{\partial y^2} S_x $, such that

\begin{equation}
    \Delta \boldsymbol{S} =\left(\frac{\partial^2 \theta}{\partial x^2}+\frac{\partial^2 \theta}{\partial y^2}\right)\left(\begin{array}{l}-S_y \\ +S_x\end{array}\right)=\Delta \theta\left(\begin{array}{c}-S_y \\ +S_x\end{array}\right)
\end{equation}

On the other hand, one has :
\begin{equation}
\begin{aligned}
& (\nabla \times \boldsymbol{S}) \times \boldsymbol{S}=(\nabla \times \boldsymbol{S})_z\left(\begin{array}{l}
-S_y \\
+S_x
\end{array}\right) \\
\end{aligned}
\end{equation}

Combining both parts leads to
\begin{equation}
\begin{aligned}
a^2 \Delta \boldsymbol{S}+2\sigma a\,(\nabla \times \boldsymbol{S}) \times \boldsymbol{S}
&=\left[a^2 \Delta \theta+2\sigma a\,(\nabla \times \boldsymbol{S})_z\right]\left(\begin{array}{c}
-S_y \\
+S_x
\end{array}\right) \\
& =\frac{\gamma}{J} \dot{\theta}\left(\begin{array}{l}
-S_y \\
+S_x
\end{array}\right) \\
& =\frac{\gamma}{J} \dot{\boldsymbol{S}} \\
&
\end{aligned}
\end{equation}

Defining $\tilde \sigma = 2\,\sigma\,a$, one obtains 

\begin{equation}
    \frac{\gamma}{J}\dot{\boldsymbol{S}} = a^2 \Delta \boldsymbol{S}+\tilde \sigma \,(\nabla \times \boldsymbol{S}) \times \boldsymbol{S}
\end{equation}
Finally, one can relax the hard constraint $|\boldsymbol{S}|=1$ and only softly enforce a preferred magnitude (here unity without loss of generality) via a Lagrange multiplier term $\alpha\,(1-\boldsymbol{S}^2)\boldsymbol{S}$:

\begin{equation}
    \frac{\gamma}{J}\dot{\boldsymbol{S}} = a^2 \Delta \boldsymbol{S}+\tilde \sigma \,(\nabla \times \boldsymbol{S}) \times \boldsymbol{S} + \alpha\,(1-\boldsymbol{S}^2)\boldsymbol{S}
\end{equation}

As $\sqrt{J/\alpha}$ physically sets the defects core size, it has to be larger than the simulation mesh spacing. 
In the simulations of the continuum model, we use $\alpha = 100$, $J=1$, and typically $L=256$. The mesh spacing is $1/256 \approx 0.004$ while $\sqrt{J/\alpha}=0.1$.

\section{Kosterlitz-Thouless Critical Temperature}
As described by Kosterlitz-Thouless theory \cite{Kosterlitz1973, Kosterlitz1974}, the $2d$XY model exhibits a transition from a quasi-ordered phase to a disordered one, due to the unbinding of topological defects, at a temperature denoted $T_{KT}$. 

At high temperatures $T>T_{KT}$, the entropy term wins and single isolated defects become energetically favourable: the proliferate and span the entire system, leading to disorder, characterised by exponentially decaying spatial correlation functions $C(r) \sim e^{-r/\xi}$, where $\xi$ is the correlation length.
At low temperatures, the only energetically favourable way for a defect to exist is to be bound with an oppositely charged defect, thus creating a pair. As the perturbation created by bounded defects gets limited to the distance between them, the system can reach quasi-long-range order, characterised by the power-law decay of spatial correlation functions $C(r) \sim r^{-\eta}$. In particular, at the transition temperature $T=T_{KT}$, renormalisation group calculations predict $\eta = 1/4$. 

On the square lattice, the value of the critical temperature $T_{KT}=0.89$ has been deeply investigated \cite{Hasenbusch2005, littTKT}. However,  the literature is rather scarce regarding the triangular lattice case. Butera and Comi presented in 1994 a high-temperature expansion and concluded $T_{KT}=1.47$ \cite{butera1994high} . Here, we directly measure the spatial correlation function $C(r)$ for different temperatures. The critical temperature $T_{KT}$ is here determined as the largest temperature for which $C(r)$ is well described by a power-law decay. We present the results in Fig.\ref{fig:TKT} for the square lattice (left panel) and for the triangular lattice (right panel). For the square lattice, we obtain $T_{KT}=0.89$, in  agreement with the reported values. For the triangular lattice, we obtain $T_{KT}=1.4$, compatible with the value $T_{KT}=1.47$ reported in \cite{butera1994high}. 

\begin{figure}[h!]
    \centering
    \includegraphics[width=0.65\linewidth]{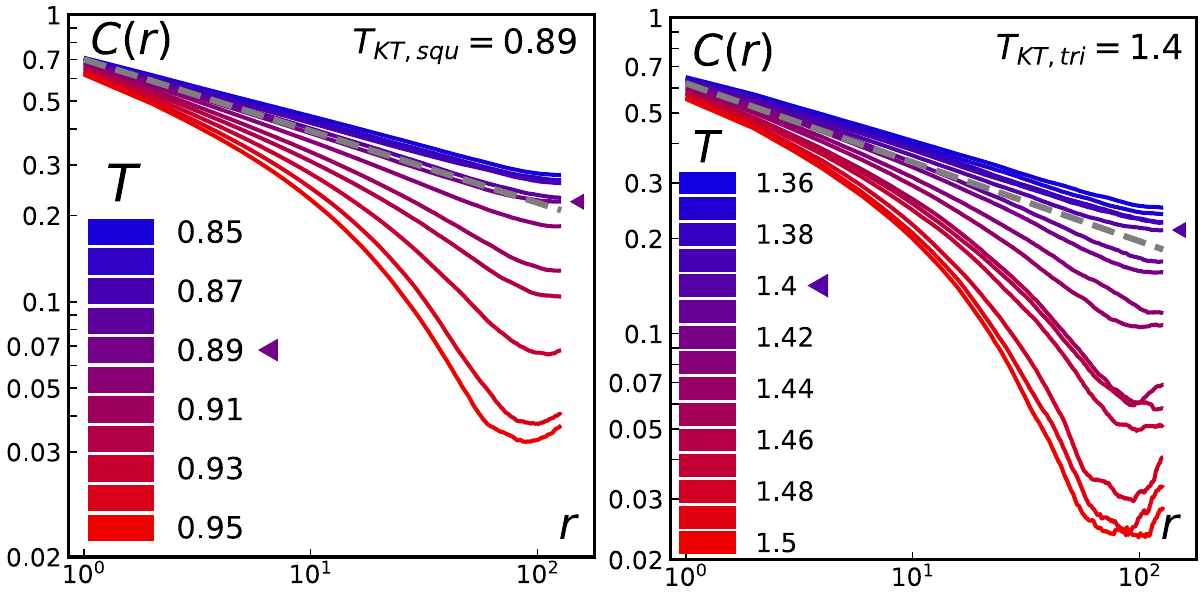}
    \caption{Spatial correlation function $C(r)$ computed for a $L=256$ system. \textbf{(left)} Square lattice, we obtain $T_{KT}\approx 0.89$ \textbf{(right)} Triangular lattice, we obtain $T_{KT}\approx 1.4$.}
    \label{fig:TKT}
\end{figure}

Let's denote by $T_4$ (resp. $T_6$) the critical temperature for the square (resp. triangular) lattice.
Interestingly, the ratio $\frac{T_6}{T_4} = \frac{1.4}{0.89 } = 1.57 \neq \frac{6}{4}$, meaning that there is more going on than a simple rescaling due to the number of neighbours (as a simple mean-field treatment would predict).

\section{Growth of the polarisation axis with $\sigma$}

Non-reciprocal interactions create large scale collective structures such as the polarisation axis of the $q=-1$ defects. The larger $\sigma$, the more influenced spins are by the neighbours they look at. This favouring large coherent structures and the growth of the polarisation axis is faster for larger $\sigma$. 

To quantify such process, one indirect but simple and informative {figure of merit} is the \emph{nematic} order parameter $Q = \frac{1}{N}\sum_j e^{2i\theta_j}$. 
We show in Fig.\ref{fig:axis_growth} the time evolution of $Q$ for a system composed of a single negative defect (centered). The initial condition is a perfectly isotropic defect $\theta(x,y) = -\text{atan}(y/x) + \mu_-$, $Q(t=0) =0$, $P(t=0) =0$. With time, the polarisation axis extends, separating two regions in which the \emph{polar} order $P$ increases. Yet, those two regions have opposite orientation, so the global $P$ remains zero throughout the process. However, the nematic order parameter $Q$ increases as the contributions from the two regions add up. The left panel of Fig.\ref{fig:axis_growth}  shows that the larger the non-reciprocal parameter $\sigma$, the faster the dynamics at small times. The right panel of Fig.\ref{fig:axis_growth} shows the same data but as a function of $\sigma t$ (in a logarithmic scale to emphasis the short time regime). 
However, the long time regime also depends on the elasticity of the $\theta$ field, and hence how thin the polarisation axis can be. The larger $\sigma$, the thinner the axis can be (because the non-reciprocal forces dominate over the elastic ones), so the system tends to be fully polarised and $Q(t\to \infty)$ increases.

Finally, the value of $\mu_-=\text{Arg}(\sum_{j=1}^3 e^{i[\theta_j +\,\text{atan}(y_j/x_j)]} )$ remains constant throughout this polarisation process. Indeed, even though the shape dramatically changes, losing its radial symmetry and almost gaining a translational symmetry, the field around a negative defect never looses its central symmetry, ie. the contributions from opposite quadrants (the origin being the defect core) remain equal and opposite. This implies that the value of $\mu_-$ (the way we defined it) does not vary.  

\begin{figure}[h!]
    \centering
    \includegraphics[width=0.65\linewidth]{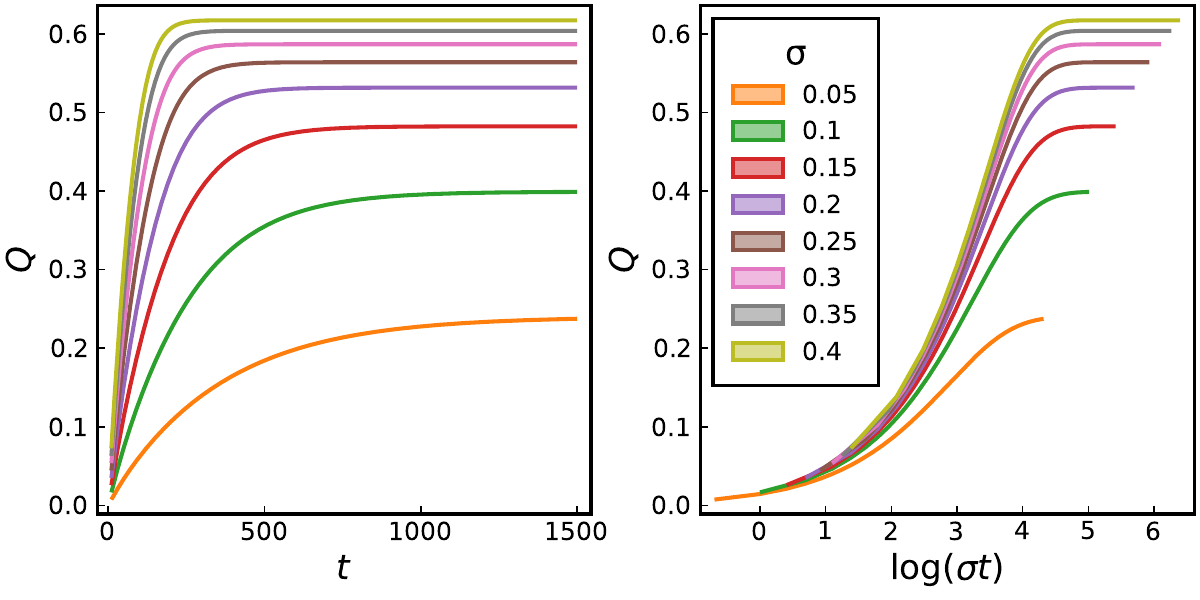}
    \caption{(\textbf{left}) Nematic order parameter $Q(t)$ of an isolated defect in a $50\times50$ system at $T=0$. Different colours correspond to different $\sigma$, as indicated in the right panel. (\textbf{right}) Same data, but against $\log(\sigma t)$ instead of $t$ . }
    \label{fig:axis_growth}
\end{figure}

\section{{Details on the $\tau = \tau_{XY}$ line in Fig. 3(h)}}

{In this section, we briefly come back to the Fig. 3(h) of the main text to discuss the conditions under which the annihilation time corresponds to the one in the reciprocal system, i.e.  $\tau = \tau_{XY}$, drawn with white dots in Fig.~\ref{fig:tau_tauXY_line}.}
\begin{figure}[h!]
    \centering
    \includegraphics[width=0.65\linewidth]{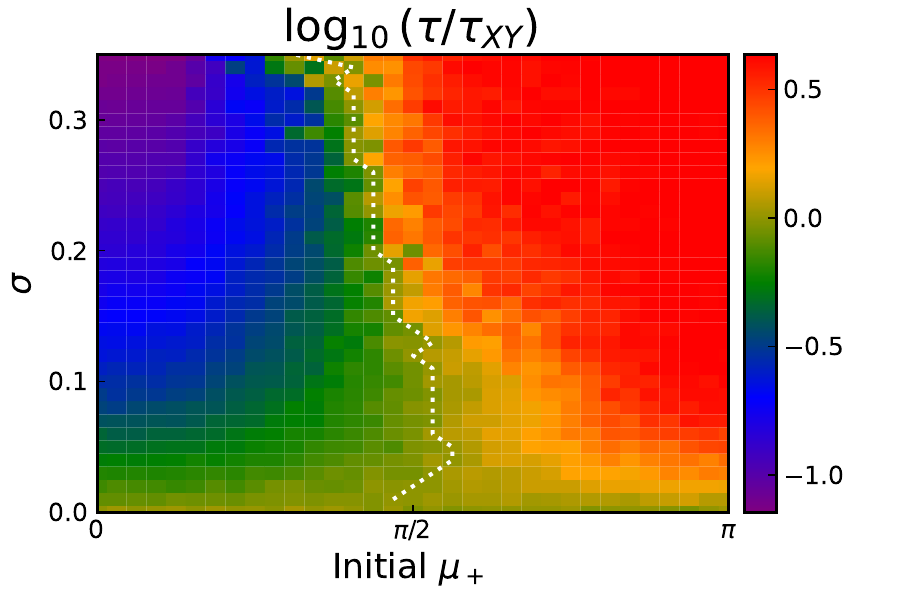}
    \caption{Same Figure as Fig. 3(h) of the main text, to which we add the white dots to visualise the $\tau = \tau_{XY}$ line. 
    Colour map of the annihilation time in log-scale 
     as a function of initial shape $0\leq\mu_+\leq\pi$ and non-reciprocity $\sigma$ (using the lattice model, $L=64, R_0 = 30$ and $T = 0.08\ T_{KT}$. }
    \label{fig:tau_tauXY_line}
\end{figure}
{This curve seems to become vertical at high enough non-reciprocity $\sigma$, around a value $\mu_+$ that happens not to be $\pi/2$. The case $\mu_+= 0$ gives an enhanced annihilation and the case $\mu_+= \pi$ gives a hindered annihilation, both with motion along the core-to-core axis. There is a priori no reason why $\mu_+= \pi/2$ should produce an annihilation time identical to that of the equilibrium XY case. Indeed, as discussed in Fig. 3(c) of the main text as well as in the Supplementary Movies 6(a,b), the general scenario for the annihilation dynamics for an initial $\mu_+ \neq 0,\pi$ is the following: the $+1$ defect twists to the sink state $\mu_+ =\pi$, the $-1$ defect grows its polarisation axis diagonally and runs along it. The motion of the two defects is no longer reciprocal nor along the core-to-core axis as in the XY case, so we don't expect the $\tau = \tau_{XY}$ line to bring deeper information. In the case of the present Fig.~\ref{fig:tau_tauXY_line}, the final value of the plateau is $\mu_+\approx 0.875\,\pi/2$ . We note that the precise location of the $\tau = \tau_{XY}$ line might depend on the temperature but does not depend on the initial distance between the defects. 
}

\section{Rewriting the model of Ref. \cite{vafa2022defect} in terms of vector calculus operators}
In their work \cite{vafa2022defect}, Vafa considered the following equation: 

\begin{equation}
\begin{aligned}
   \partial_t p=\mathcal{I}(p)&=-\frac{\delta \mathcal{F}(\{p\})}{\delta \bar{p}}+\lambda \mathcal{I}_\lambda(p) \\
&=4 \partial \bar{\partial} p+2 \epsilon^{-2}\left(1-|p|^2\right) + \lambda(p \partial+\bar{p} \bar{\partial})p
\end{aligned}
\end{equation}

They defined

\begin{equation*}
\begin{aligned}
p & =p_x+i p_y \\
\bar{p} & =p_x-i p_y \\
\partial & =\frac{1}{2}\left(\partial_x-i \partial_y\right) \\
\bar{\partial} & =\frac{1}{2}\left(\partial_x+i \partial_y\right) \\
\end{aligned}
\end{equation*}

Therefore

\begin{equation*}
\begin{aligned}
4 \partial \bar{\partial} & =\left(\partial_x-i \partial_y\right)\left(\partial_x+i \partial_y\right)=\partial_x^2+\partial_y^2=\Delta \\
2 p {\partial} & =\left(p_x+i p_y\right)\left(\partial x-i \partial_y\right) \\
& =p_x \partial_x+p_y \partial_y+i\left(p_y \partial_x-p_x \partial_y\right) \\
2 \overline{p} \bar{\partial} & =\left(p_x-i p_y\right)\left(\partial_x+i \partial_y\right) \\
& =p_x \partial_x+p_y \partial_y-i\left(p_y \partial_x-p_x \partial_y\right)
\end{aligned}
\end{equation*}

So $(p \partial+\bar{p} \bar{\partial})p=(\vec{P} \cdot \nabla) \vec{P} \quad$ where $\quad \vec{P}=
\begin{pmatrix}
           p_x \\
        p_y 
         \end{pmatrix}$ .

\section{Supplementary discussion on the work of Loos, Klapp and Martynec (Phys. Rev. Lett. , 2023)}

In their article \emph{Long-Range Order and Directional Defect Propagation in the Nonreciprocal XY Model with Vision Cone Interactions} (2023) \cite{loos2023long}, Loos, Klapp and Martynec introduced a similar $2d$XY model with non-reciprocal couplings. In this section, we discuss the differences between both approaches.

\subsection{{Interplay between vision cones and the discrete nature of the lattice}}

In Ref.\cite{loos2023long}, a spin $j$ is considered to be in the neighbourhood of a spin $i$ if and only if $j$ is in the vision cone of $i$. 
They use a sharp vision cone kernel, as described in orange in Fig.1 of the main text, such that $j$ is an interacting  neighbour of $i$ if and only if the angle between the orientation $\theta_i$ of the spin $i$ and the location $u_j$ of the spin $j$ is less than half of the vision cone $\Theta$. 

{The main consequence of the interplay between a sharp kernel and a discrete underlying lattice is the generic \ylann{sudden} change in the number of neighbours as the system evolves in time, as we illustrate here on a triangular lattice ($z=6$ is thus the coordination number of the lattice). }
This aspect is discussed in depth in \cite{loos2023long}. The case $\Theta=2\pi/z, \ n=1,...,z$, illustrated in Fig.\ref{fig:issues_vision_cone}(a,b), is helpful to understand this {feature}. On the triangular lattice, if the orientation of a spin $i$ is $\theta_i = \pi/2$, $i$ has 4 neighbours in its vision cone. If its orientation fluctuates by an infinitesimal amount, it immediately looses one neighbour. The same happens for all values of $\Theta$,  as we sketch in Fig.~\ref{fig:issues_vision_cone}(c,d) for an arbitrary vision cone amplitude. The consequence of this brutal change in the number of neighbours is important. If one works with a Monte-Carlo type of dynamics, since the energy is extensive (as opposed to intensive if normalised by the number of contributing neighbours) there exist $z$ regions energetically favourable
imposed by the symmetry of the underlying lattice. 
This strongly promotes a local alignment with the symmetry of the lattice. 

If instead one works with a Langevin approach based on extensive (i.e. additive) forces, a similar feature remains. We illustrate in Fig.~\ref{fig:issues_vision_cone}(e) a configuration where the blue spin (in the center) has only one {neighbour} within its vision cone, while the green spin (on the right) has two neighbours, among which the one in blue. The Langevin equation of motion looks like $\dot \theta_i = \sum_j \sin(\theta_j-\theta_i) +\sqrt{2T}\nu_i$. For the blue spin, the sum only contains one contribution while for the green spin, it contains 2 terms so it is potentially twice as big. However, the temperature $T$ of the thermal bath is constant. 
The ratio between the thermal angular diffusion and the interacting force thus becomes space and time dependent.

To avoid these consequences, one could imagine normalising the energy (or the force in a Langevin-like approach) by the (time-dependent)  number of neighbours within the vision cone. However, this introduces another source of non-reciprocity, as  the force of green on blue is now twice the force of the blue on green, in absolute value. The introduction of this new source of non-reciprocity has been documented in \cite{chepizhko2021revisiting}. 

\begin{figure}[h!]
    \centering
    \includegraphics[width=0.5\linewidth]{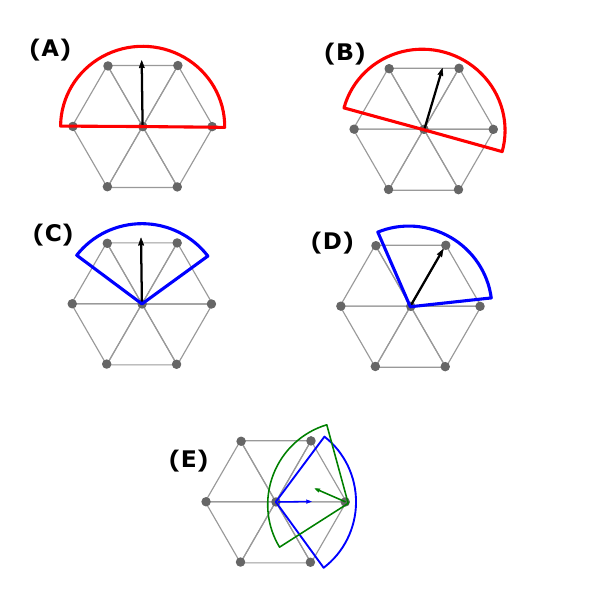}   
    \caption{ 
   \textbf{(A, B, C, D)} A spin can brutally lose or gain one neighbour within its sharp vision cone during the dynamics. \textbf{(E)} The effective temperature felt by two neighbouring spins can be different if their number of neighbours is different.
    }
    \label{fig:issues_vision_cone}
\end{figure}

\ylann{These sharp features are a consequence of using a sharp vision cone kernel on a discrete lattice. 
 However, the interplay with the symmetry of the underlying lattice is a generic feature of such vision cone models (sharp and smooth) on discrete lattices. As such, even if we have not quantified the effective alignment and its impact at larger scales (on defects or at the system scale) in the present work, an interplay with the discrete lattice, providing biased directions for the global order parameter, is also expected.  

Indeed, since the kernel $g(\varphi) = \exp(\sigma\,\cos \varphi)$ is continuous and differentiable, a small change in $\varphi$ can only generate a small change in $g$. Such a feature of vision cone models on regular lattices has been recently discussed in \cite{bandini2024xy, popli2025don}.
 }

A relevant measure is the sum of the angular weights $\mathcal{G}_i  \equiv \sum_{j=1}^z g\left(\varphi_{i j}\right)$, as it relates to the maximum force a spin can feel. In particular, its dependence on $\theta_i$ is related to the discretisation effects; its dependence on $\sigma$ allows to understand the relative importance of the two terms in the equation of motion: alignment and thermal noise.
We have seen above that, in the sharp vision cone case, both a change in orientation $\theta_i$ or in the vision cone amplitude $\Theta$, can result in a sudden gain of one neighbour, which translates into a sudden increase (+1) of $\mathcal{G}_i$ . 
In the soft vision cone case, we want to study the dependence of $\mathcal{G}_i$ on $\sigma$ and $\theta_i$. To do so, we consider the square lattice for two reasons: (i) the calculations are simpler (ii)  discretisation effects in the triangular lattice are less pronounced than for the square lattice; we thus consider the worst-case scenario. The sum over the angular weights gives:
\begin{equation}
\begin{aligned}
\mathcal{G}_i & \equiv \sum_{j=1}^4 g\left(\varphi_{i j}\right) \\
& =\sum_{j=1}^4 e^{\sigma \cos \left(\theta_i-u_{j}\right)} \\
& =\sum_{j=1}^4 e^{\sigma \cos \left(\theta_i-(j-1) \frac{\pi}{2}\right)} \\
& =e^{\sigma \cos \theta_i}+e^{\sigma \sin \theta_i}+e^{-\sigma \cos \theta_i}+e^{-\sigma \sin \theta_i } \\
& =2 \cosh \left(\sigma \cos \theta_i\right)+2 \cosh \left(\sigma \sin \theta_i\right)
\end{aligned}
\end{equation}
Since $|\cos x|\leq1$ and $|\sin x|\leq1$ , for small $\sigma$ one can Taylor expand the cosh:
\begin{equation}
\begin{aligned}
\mathcal{G}_i & = 2+\left(\sigma \cos \theta_i\right)^2+\frac{1}{12}\left(\sigma \cos \theta_i\right)^4+\mathcal{O}\left(\sigma^6\right) \\
& +2+\left(\sigma \sin \theta_i\right)^2+\frac{1}{12}\left(\sigma \cos \theta_i\right)^4+\mathcal{O}\left(\sigma^6\right) \\
& =4+\sigma^2+\sigma^4\frac{\cos ^4 \theta_i+\sin ^4 \theta_i}{12}+\mathcal{O}\left(\sigma^6\right)
\end{aligned}
\end{equation}

We learn that $\mathcal{G}_i $ only varies with the orientation $\theta_i$ to \emph{fourth} order in $\sigma$. 
For $\sigma=0.5$ , it implies that $\mathcal{G}_i(\theta) $ varies at most by 0.06 \% from it maximum value $\mathcal{G}_i(\theta=n\pi/2), \ n=1,...,z$ (to be compared to the 25 \% of the vision cone model when a spin passes from 4 to 3 neighbours). 
Indeed, when $\theta = \pi/4$, the mode of the kernel falls in between 2 neighbours, and

\begin{equation}
\begin{aligned}
\frac{\mathcal{G}_i(\theta=0;\sigma=0.5)-\mathcal{G}_i(\theta=\pi/4;\sigma=0.5)}{\mathcal{G}_i(\theta=0;\sigma=0.5)} & \\
= \frac{4.2552- 4.2526}{4.2552} = 0.0006
\end{aligned}
\end{equation}

Therefore, the soft vision kernel we propose should indeed strongly attenuates the effects stemming from the underlying discrete lattice for the typical range of $\sigma \in [0,1/2]$ used in this work. 

\subsection{Small vision cones and percolation threshold}
Most of the results presented in this work are  for values of $\sigma \leq 0.5$. We pushed up to $\sigma=1$ for Fig.3(e) of the main text. 
However, one cannot increase $\sigma$ up to arbitrarily large values. Indeed, the larger $\sigma$ gets, the more peaked the kernel $g(x) = \exp(\sigma \cos x)$ becomes. If almost all the mass of the distribution falls between two neighbours, the spin in fact ``sees" no neighbour at all, and its dynamics is thus only driven by white noise. 
In this Letter, we work far from this  regime. For instance, for $\sigma=0.5$, the deviation from an isotropic kernel is moderate:  $g(0)=e^{0.5} \approx 1.65$ and the smaller coupling value is $g(\pi)=e^{-0.5} \approx 0.6$, far from negligible.
Such pathological limit $\sigma \to \infty$ is not a specificity of our model. Loos et al. \cite{loos2023long}, for their sharp vision cone model, report in the same spirit that for vision cones smaller than $\pi/3$ on the triangular lattice ($\pi/2$ on the square lattice), the percolation of the interaction network is no longer guaranteed.

\subsection{Equivalence between sharp and soft vision cones}
\begin{figure}[h!]
    \centering
    \includegraphics[width=0.7\linewidth]{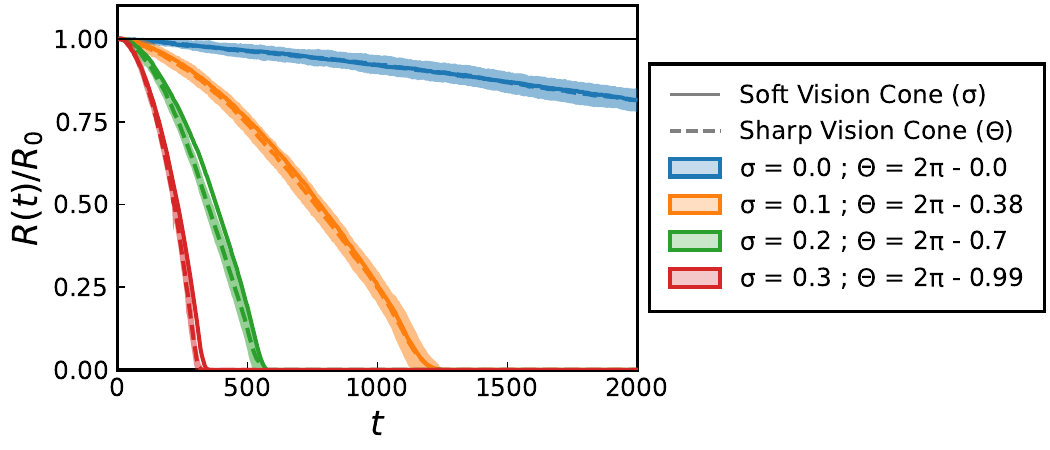}
    \caption{
    Distance separating two defects over time in the enhanced annihilation configuration (the $+1$ defect being a source $\mu_+=0$. The annihilation dynamics of both models quantitatively match if the sharp vision cone aperture is taken from Eq.(\ref{eq:equivalence_short_sharp_vision_cones}). Results for the sharp vision cone are averaged over $R=50$ independent realisations of the thermal noise. The standard deviations are shown with the ribbons.}
    \label{fig:equiv_vision_cones}
\end{figure}
For soft vision cones, we use the non-reciprocal parameter $\sigma\geq0$ used throughout the main manuscript. For  sharp vision cones, we use a vision cone $\Theta \leq 2\pi$, centered on the phase $\theta$ of the spins. \\
Sharp and soft vision kernels ($g_\text{sharp}$ and $g_\text{soft}$) have a common property: the total weight in the head-side hemisphere ($\varphi \in [-\pi/2, \pi/2]$) is greater than the total weight of the tail-side hemisphere ($\varphi \in [\pi/2, 3\pi/2]$). We define the ratio $f(\sigma)$ between those as 
\begin{equation}
    f(\sigma) = \int\limits_{-\pi/2}^{\pi/2} d\varphi \, g(\varphi)\bigg/\int\limits_{\pi/2}^{3\pi/2} d\varphi \, g(\varphi)
\end{equation}

For a reciprocal (=isotropic) kernel $g(\varphi) = 1$, then $f(\sigma)=1$;  while $f(\sigma)>1$ for non-reciprocal kernels.  \\

\subsubsection{Soft vision cones}
For soft vision cones $g_\text{soft}(\varphi) = \exp(\sigma \,\cos \varphi)$, one obtains
\begin{equation}
    f_\text{soft}(\sigma) = \frac{\pi(L_0(\sigma) + I_0(\sigma))}{\pi(L_0(\sigma) - I_0(\sigma))}
\end{equation}
where {$I_0$} is the modified Bessel function of the first kind and { $L_0$} is the modified Struve function. 
For small $\sigma$, one can Taylor expand those functions up to second order to obtain
\begin{equation}
    f_\text{soft}(\sigma) \approx \frac{\pi + 2\sigma + \pi \sigma^2/4}{\pi - 2\sigma + \pi \sigma^2/4} \ .
\end{equation}

\subsubsection{Sharp vision cones}
For {sharp} vision cones $$g_\text{{sharp}}(\varphi) = 
\begin{cases}
1 \text{ if } |\varphi| \leq \Theta/2 \\
0 \text{ otherwise}  
\end{cases}\ ,$$
one obtains
\begin{equation}
    f_\text{sharp}(\Theta) = \frac{\pi}{\Theta-\pi} \ .
\end{equation}

Finally, equating the head-tail asymmetry ratios ($ f_\text{sharp} =  f_\text{soft}$)  gives an equivalence relation between $\sigma$ and $\Theta$:

\begin{equation}
    \Theta(\sigma) = 2\pi - \frac{16\pi\sigma}{\pi(4+\sigma^2) + 8\sigma }
    \label{eq:equivalence_short_sharp_vision_cones}
\end{equation}
which {allows to compare} the dynamics of defects found in both versions of the model, as shown in Fig.\ref{fig:equiv_vision_cones}.

\subsection{Equation of motion stemming from the XY energy modulated by the kernel}
We now compare our Langevin equation of motion to the equation of motion one would obtain from the relaxation driven by an  energy function
\begin{equation}
E=-\sum_{i, j} g\left(\varphi_{i j}\right) \cos \left(\theta_j-\theta_i\right)
\end{equation}
If $g(x) = \exp(\sigma \, \cos x)$, the corresponding equations of motions would be (with $\Delta \theta_{ij} \equiv \theta_j - \theta_i$ and $\varphi_{ij} \equiv \theta_i - u_j$) : 
\begin{equation}
\begin{aligned}
&\dot{\theta}_i  =-\frac{\partial E}{\partial \theta_i} \\
 &=\sum_{j} \frac{\partial}{\partial \theta_i}\left(g\left(\varphi_{i j}\right) \cos \left(\Delta \theta_{ij}\right)\right) \\
 &=\sum_j - g\left(\varphi_{i j}\right) \sin \left(\Delta \theta_{ij}\right) \underbrace{\frac{\partial \Delta \theta_{i j}}{\partial \theta_i}}_{=-1}+\cos \left(\Delta \theta_{ij}\right) \frac{\partial g\left(\varphi_{i j}\right)}{\partial \theta_i} \\
 &=\sum_j g\left(\varphi_{i j}\right) \sin \left(\Delta \theta_{ij}\right)+\cos \left(\Delta \theta_{ij}\right)\left(-\sigma \sin \varphi_{i j}\right) \underbrace{\frac{\partial \varphi_{i j}}{\partial \theta_i}}_{=1} g\left(\varphi_{i j}\right) \\
 &=\sum_j g\left(\varphi_{i j}\right)\left[\underbrace{\sin \left(\Delta \theta_{ij}\right)}_{\text{attraction}}-\underbrace{\sigma \sin \left(\varphi_{i j}\right) \cdot \cos \left(\Delta \theta_{ij}\right)}_{\text{(side) repulsion !}}\right]
\end{aligned}
\label{eq:extra_term}
\end{equation}
The first term $\sin\left(\theta_j-\theta_i\right)$ corresponds to the alignment term in our Langevin description. 
The additional second term  in $\cos \left(\theta_j-\theta_i\right)$ favours orthogonal spin configurations but, due to the $\sin \left(\varphi_{i j}\right)$ coefficient, this repulsion mainly takes place on the sides of the spin $i$, ``ahead" being in the direction of $\theta_i$. The amplitude of this second term, however, is \emph{a priori} smaller than the first term, since in this work we used $\sigma\leq1$ and since $|\sin \left(\varphi_{i j}\right)|\leq1$.
Yet, the impact of this additional term, though only to second order, favours the stabilisation of the sources $\mu_+ \approx 0$, a configuration in which spins are indeed perpendicular. 

\begin{figure}[h]
    \centering
    \includegraphics[width=0.5\linewidth]{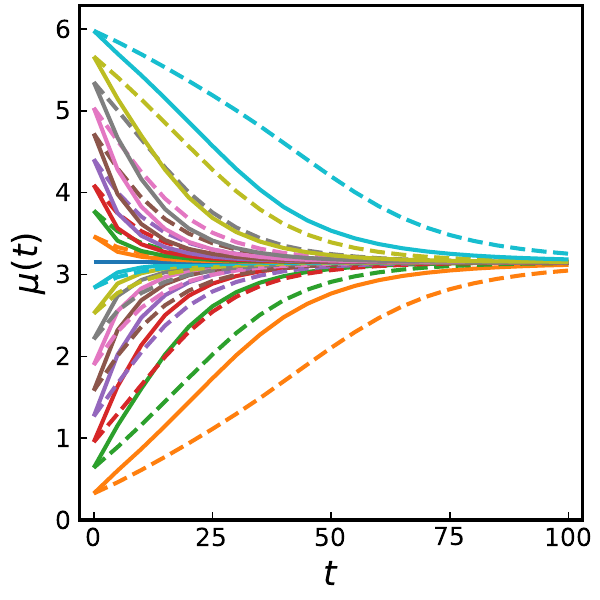}
    \caption{In the spirit of Fig.2(g) of the main text, we monitor the shape $\mu_+(t)$ of a positive defect over time, for different initial shapes $\mu_+(0)$ (different colours). The solid lines correspond to our Langevin-based model, with only the first alignment term. The dashed lines correspond to equation Eq.(\ref{eq:extra_term}), with the additional second term.}
    \label{fig:extra_term}
\end{figure}

We report the effect of this additional term on the twist of positive defects in Fig.\ref{fig:extra_term}. 
As stated, the $\cos \left(\theta_j-\theta_i\right)$ term only has a small amplitude and thus does not change the stable configuration which remains the sink state $\mu_+ = \pi$. However, the decay to this stable state is slower, especially for initial configurations close to sources $\mu_+ = 0 = 2\pi$ (cf. orange or cyan curves). 
This could have large-scale consequences, as the sources are the only defects involved in the enhanced  annihilation process. If they are slightly stabilised by this repulsive term, the enhanced annihilation process could be even more dramatic and impact the overall coarsening dynamics. 

\end{document}